\documentclass[aps,prl,10pt,twocolumn,amsmath,amssymb,longbibliography,superscriptaddress,nofootinbib]{revtex4-2}
\usepackage{amsmath,amssymb,amsfonts} 	
\usepackage{graphicx}
\usepackage{hhline}
\usepackage[latin1]{inputenc}
\usepackage{subfigure}
\usepackage[normalem]{ulem}
\usepackage{xcolor}

\begin{document}
\title{Stoner contributions to the magnon equation of motion}

\author{Thorbj\o rn Skovhus}
\email{torbjorn.skovhus@physics.uu.se}
\affiliation{Department of Physics and Astronomy, Uppsala University, 
Box 524, 
751 20 Uppsala, Sweden}
\author{Patrik Thunstr\"{o}m}
\affiliation{Department of Physics and Astronomy, Uppsala University, 
Box 524, 
751 20 Uppsala, Sweden}

\begin{abstract}

From linear response theory, we derive the equation of motion for Landau-damped magnon quasi-particles in absence of spin-orbit coupling. The derivation is based on a minimal set of assumptions, namely that the transverse magnetic susceptibility is diagonalized by one collective eigenmode per magnetic atom in the unit cell, and that the spectrum of each collective eigenmode is dominated by a single magnon resonance. 
The resulting equation of motion leads to a natural definition of the exchange interaction as the restoring force acting on the collective mode magnetization near equilibrium. In addition to magnetic exchange, the magnon equation of motion also contains a damping term and a dispersive quasi-particle mass, both originating from the electron-magnon coupling. The effect of these terms is illustrated for a prototypical itinerant ferromagnet, where they redshift the magnon dispersion and reduce the magnon lifetime as the magnon enters the Stoner continuum. 
By further assuming wave vector independence of the collective magnon subspace, the magnon equation of motion becomes atomistic with well defined magnetic sites. For real materials, one can approximate the magnon coupling constants using an arbitrary level of band theory for the dynamic transverse magnetic susceptibility.

\end{abstract}
\maketitle


Traditionally, the study of magnetic excitations is based on a mapping of the electronic structure to a Heisenberg model of atomic magnetic moments \cite{Liechtenstein1987,Halilov1998,Grotheer2001,Antropov2003,Bruno2003,Szilva2023,Solovyev2024}. 
The mapping relies on a full decoupling of the (slow) magnetic degrees of freedom from the (fast) electronic degrees of freedom, and results in a description of collective magnetic excitations---magnons---which is governed solely by exchange. 
In itinerant magnets, however, magnons are Landau damped due to an explicit coupling to the continuum of spin-flipped electron-hole pairs \cite{Moriya1985}, and even at the level of an adiabatic decoupling, the spin dynamics involve a dispersive "Berry curvature" in addition to the exchange interaction \cite{Niu1998,Niu1999}.
Instead of decoupling electrons and magnons, one can extract the full spectrum of magnetic excitations from a first principles calculation of the dynamic susceptibility \cite{Savrasov1998,Aryasetiawan1999,Karlsson2000,SasIoglu2010,Buczek2011b,Lounis2011,Rousseau2012,Muller2016,Cao2017,Singh2019,Okumura2019,Tancogne-Dejean2020,Friedrich2020,Skovhus2021,Liu2023,Skovhus2026a}. Although such methods have matured significantly, they are expensive, and calculations of derived physical properties quickly become impractical.
In this Letter, we therefore seek a middle ground: a magnon equation of motion which includes the coupling to the Stoner continuum, derived by making a minimal set of assumptions to the transverse spin dynamics near thermal equilibrium \cite{Qian2002}.


From the perspective of linear response, the dynamics of an electronic system can be characterized by the time evolution of its four-component electron density, $\hat{n}^\mu(\mathbf{r})=\sum_{s,s'}\sigma^\mu_{ss'}\hat{\psi}_s^\dagger(\mathbf{r})\hat{\psi}_{s'}(\mathbf{r})$ with $\mu\in\{0,x,y,z\}$ and $\sigma^0=\mathbb{I}_2$. The construction is as follows. Initially, the system is assumed to be in thermal equilibrium, $\langle\hat{n}^\mu(\mathbf{r},t)\rangle=\langle\hat{n}^\mu(\mathbf{r})\rangle_0$ for $t\leq t_0<0$. Then, for a finite duration of time $-t_0$, the system is weakly perturbed by an external electromagnetic field. This induces a change in the density, which to linear order in the field strength is governed by the dynamic susceptibility $\chi^{\mu\nu}$,
\begin{equation}
    \delta n^\mu(\mathbf{r}, t) = \sum_\nu \int_{-\infty}^\infty dt' \int d\mathbf{r}'\, \chi^{\mu\nu}(\mathbf{r}, \mathbf{r}', t-t') W_\mathrm{ext}^\nu(\mathbf{r}', t')
    \label{eq:linear response relation}
\end{equation}
with $W_\mathrm{ext}^\mu = \left(-e \phi_\mathrm{ext}, \mu_\mathrm{B} \mathbf{B}_\mathrm{ext} \right)$. 
By construction, $W_\mathrm{ext}^\mu(\mathbf{r}, t)=0$ for $t\leq t_0$ and $t\geq 0$, meaning that the response can be temporally divided into the driven and free response respectively, $\delta n^\mu(\mathbf{r}, t)=\delta n_\mathrm{d}^\mu(\mathbf{r}, t)+\delta n_\mathrm{f}^\mu(\mathbf{r}, t)$, where $\delta n_\mathrm{d}^\mu(\mathbf{r}, t)=\delta n^\mu(\mathbf{r}, t)\left[1-\theta(t)\right]$ and  $\delta n_\mathrm{f}^\mu(\mathbf{r}, t)=\delta n^\mu(\mathbf{r}, t)\theta(t)$. Inverting the response relation \eqref{eq:linear response relation} in the frequency domain,
\begin{equation}
    \sum_\nu \int d\mathbf{r}'\, \chi^{-1}_{\mu\nu}(\mathbf{r}, \mathbf{r}', \omega) \delta n_\mathrm{f}^\nu(\mathbf{r}', \omega) = W_\mathrm{eff}^\mu(\mathbf{r}, \omega),
    \label{eq:generalized equation of motion}
\end{equation}
one can directly infer the equation of motion for $\delta n_\mathrm{f}^\mu(\mathbf{r}, t)$, thus characterizing the freely propagating many-body quantum system near thermal equilibrium. 
Taylor expanding  $\chi^{-1}_{\mu\nu}$ around the static $\omega\rightarrow 0$ limit, frequency orders $(-i\omega)^n$ become $n$'th order time derivatives $(\partial/\partial t)^n$ in the integro-differential equation for $\delta n_\mathrm{f}^\mu(\mathbf{r}, t)$, while 
\begin{equation}
    W_\mathrm{eff}^\mu(\mathbf{r}, \omega) = W_\mathrm{ext}^\mu(\mathbf{r}, \omega) - \sum_\nu \int d\mathbf{r}'\, \chi^{-1}_{\mu\nu}(\mathbf{r}, \mathbf{r}', \omega) \delta n_\mathrm{d}^\nu(\mathbf{r}', \omega),
\end{equation}
determines the initial conditions.

For transverse spin dynamics in particular, Qian and Vignale \cite{Qian2002} have proposed to truncate the expansion at linear order, $\chi^{-1}_{\mu\nu}(\mathbf{r}, \mathbf{r}', \omega) \simeq \chi^{-1}_{\mu\nu}(\mathbf{r}, \mathbf{r}')+i\omega\, \tilde{\Omega}_{\mu\nu}(\mathbf{r}, \mathbf{r}')$, where $\chi^{-1}_{\mu\nu}(\mathbf{r}, \mathbf{r}')=\chi^{-1}_{\mu\nu}(\mathbf{r}, \mathbf{r}', \omega=0)$ and $\tilde{\Omega}_{\mu\nu}(\mathbf{r}, \mathbf{r}')=\lim_{\omega\rightarrow0}\partial\, \mathrm{Im}\:\chi^{-1}_{\mu\nu}(\mathbf{r}, \mathbf{r}', \omega)/\partial\omega$ (the real part is even in $\omega$). This leads to the equation of motion
\begin{align}
    \sum_\nu \int d\mathbf{r}'\, \tilde{\Omega}_{\mu\nu}(\mathbf{r}, \mathbf{r}') &\frac{\partial}{\partial t}\delta n_\mathrm{f}^\nu(\mathbf{r}', t) 
    \nonumber \\
    = &\sum_\nu \int d\mathbf{r}'\, \chi^{-1}_{\mu\nu}(\mathbf{r}, \mathbf{r}') \delta n_\mathrm{f}^\nu(\mathbf{r}', t),
    \label{eq:four-component qian-vignale}
\end{align}
%
with initial conditions $\sum_\nu \int d\mathbf{r}'\, \tilde{\Omega}_{\mu\nu}(\mathbf{r}, \mathbf{r}') \delta n_\mathrm{f}^\nu(\mathbf{r}', t=0^+) = -W_\mathrm{eff}^\mu(\mathbf{r}, \omega=0)$. 
In the most general case, Eq. \eqref{eq:four-component qian-vignale} couples all four components of the electron density, resulting in solutions of mixed spin character. For simplicity, we will therefore focus on collinear magnetic systems absent of spin-orbit coupling. 
Here, the transverse $\alpha,\beta\in\{x,y\}$ components decouple from the longitudinal, and
\begin{equation}
    \begin{pmatrix}
        \chi^{xx} & \chi^{xy} \\
        \chi^{yx} & \chi^{yy}
    \end{pmatrix}
    =
    \begin{pmatrix}
        \chi^{+-} + \chi^{-+} & i\chi^{+-} - i\chi^{-+} \\
        -i\chi^{+-} + i\chi^{-+} & \chi^{+-} + \chi^{-+}
    \end{pmatrix}
\end{equation}
is diagonalized by the circular coordinate susceptibilities
\begin{equation}
    \chi^{+-}(\mathbf{r},\mathbf{r}',t-t')=-\frac{i}{\hbar}\theta(t-t') \left\langle \left[ \hat{n}_0^+(\mathbf{r},t), \hat{n}_0^-(\mathbf{r}',t') \right] \right\rangle_0,
\end{equation}
with $\hat{n}^+(\mathbf{r})=\hat{\psi}_\uparrow^\dagger(\mathbf{r})\hat{\psi}_\downarrow(\mathbf{r})$, $\hat{n}^-(\mathbf{r})=\hat{\psi}_\downarrow^\dagger(\mathbf{r})\hat{\psi}_\uparrow(\mathbf{r})$, and $\chi^{+-}(\mathbf{r},\mathbf{r}',\omega)=\chi^{-+}(\mathbf{r},\mathbf{r}',-\omega)^*$, see e.g. Ref. \cite{Skovhus2021}. 
As a result, one can fully infer the transverse spin dynamics through inversion of $\chi^{+-}$,
\begin{equation}
    \int d\mathbf{r}_1\,\chi^{-1}_{-+}(\mathbf{r},\mathbf{r}_1,\omega)\, \chi^{+-}(\mathbf{r}_1,\mathbf{r}',\omega) = \delta(\mathbf{r}-\mathbf{r}'),
\end{equation}
where both $\chi^{+-}$ and $\chi^{-1}_{-+}$ are symmetric in the spatial entries due to Onsager reciprocity, $\chi^{+-}(\mathbf{r},\mathbf{r}',\omega)=\chi^{+-}(\mathbf{r}',\mathbf{r},\omega)$ \cite{Skovhus2022}. The inverse static susceptibility is given by $\chi^{-1}_{\alpha\beta}(\mathbf{r}, \mathbf{r}')=\delta_{\alpha\beta}\chi^{-1}(\mathbf{r},\mathbf{r}')$ with
\begin{equation}
    \chi^{-1}(\mathbf{r},\mathbf{r}')\equiv \lim_{\omega\rightarrow0}\frac{\chi^{-1}_{-+}(\mathbf{r},\mathbf{r}',\omega)}{2}, 
\end{equation}
while $\tilde{\Omega}_{\alpha\beta}(\mathbf{r},\mathbf{r}') = \varepsilon_{\alpha\beta} \Omega(\mathbf{r}, \mathbf{r}') +  \delta_{\alpha\beta}\gamma(\mathbf{r}, \mathbf{r}')$, where $\varepsilon_{\alpha\beta}$ is the two-dimensional Levi-Civita symbol,
\begin{equation}
    \Omega(\mathbf{r},\mathbf{r}') = \mathrm{Re}\:\tilde{\Omega}(\mathbf{r},\mathbf{r}'),
    \quad
    \gamma(\mathbf{r},\mathbf{r}') = \mathrm{Im}\:\tilde{\Omega}(\mathbf{r},\mathbf{r}'),
\end{equation}
and
\begin{equation}
    \tilde{\Omega}(\mathbf{r},\mathbf{r}') \equiv \lim_{\omega\rightarrow0}\frac{\partial}{\partial \omega} \frac{\chi^{-1}_{-+}(\mathbf{r}, \mathbf{r}', \omega)}{2}.
\end{equation}
In total, the transverse block of Eq. \eqref{eq:four-component qian-vignale} simplifies to
\begin{align}
    \int d\mathbf{r}'\, \big[\varepsilon_{\alpha\beta} \Omega(\mathbf{r}, \mathbf{r}')  &+ \delta_{\alpha\beta}\gamma(\mathbf{r}, \mathbf{r}')\big] \dot{m}^\beta(\mathbf{r}', t)
    \nonumber \\
    &= \int d\mathbf{r}'\, \chi^{-1}(\mathbf{r}, \mathbf{r}') m^\alpha(\mathbf{r}', t),
    \label{eq:transverse qian-vignale}
\end{align}
where $m^\alpha(\mathbf{r},t)=\delta n_\mathrm{f}^\alpha(\mathbf{r}, t)$ denotes the induced transverse magnetization and summation over the repeated $\beta$ index is implied.
The equation of motion \eqref{eq:transverse qian-vignale} generalizes the linearized Landau-Lifshitz-Gilbert equation to the continuum, and for systems with a band gap (where $\gamma(\mathbf{r},\mathbf{r}')=0$) one recovers the adiabatic equation of motion from Niu and Kleinman \cite{Niu1998}, thanks to the identification of the $\Omega$-matrix with the "Berry curvature" \cite{Qian2002}
\begin{equation}
    \Omega(\mathbf{r},\mathbf{r}')=2\hbar\, \mathrm{Im} \left\langle \frac{\partial \psi[n,\mathbf{m}]}{\partial n^x(\mathbf{r})} \right|\left. \frac{\partial \psi[n,\mathbf{m}]}{\partial n^y(\mathbf{r}')} \right\rangle.
\end{equation}

Formally, the transverse spin dynamics of Eq. \eqref{eq:transverse qian-vignale} are exact up to one central assumption; that $\chi^{-1}_{-+}(\mathbf{r},\mathbf{r}',\omega)$ is a linear function of frequency. Whereas this might be an appropriate approximation for some spatial degrees of freedom, it is difficult to justify in general. 
In addition, it is quite impractical that Eq. \eqref{eq:transverse qian-vignale} retains the full (infinite dimensional) spatial structure of the problem if we mainly are concerned with the slow (low energy) dynamics of the magnons.
In this work, we focus therefore specifically on the subset of $\chi^{+-}$ eigenmodes to which the magnons supply spectral weight and avoid an expansion of $\chi^{-1}_{-+}(\mathbf{r},\mathbf{r}',\omega)$ as a whole. Namely, we assume that the transverse magnetic susceptibility for periodic solids,
\begin{equation}
    \chi^{+-}(\mathbf{r},\mathbf{r}',\omega) = \frac{1}{N_q}\sum_\mathbf{q}\sum_{i,j} e^{i\mathbf{q}\cdot(\mathbf{r}-\mathbf{r}')}f_i(\mathbf{r})\chi_{ij}^{+-}(\mathbf{q},\omega)f_j^*(\mathbf{r}'),
\end{equation}
is diagonalized by a set of $N$ collective mode vectors $|v_n(\mathbf{q})\rangle$ for each wave vector $\mathbf{q}$,
\begin{equation}
    \chi^{+-}(\mathbf{q},\omega) |v_n(\mathbf{q})\rangle \simeq \chi_n^{+-}(\mathbf{q},\omega) |v_n(\mathbf{q})\rangle,
    \label{eq:collective mode diagonalization}
\end{equation}
where $N$ corresponds to the number of magnetic atoms in the unit cell, $\chi^{+-}(\mathbf{q},\omega)$ denotes the matrix representation of $\chi_{ij}^{+-}(\mathbf{q},\omega)$, and $f_i(\mathbf{r})=f_i(\mathbf{r}+\mathbf{R})$ yields an arbitrary orthonormal basis for the periodic degrees of freedom. The approximation \eqref{eq:collective mode diagonalization} essentially asserts a frequency-independent distinguishability between collective and single-particle modes, which turns out to be very well satisfied in elementary ferromagnets such as Fe, Ni and Co \cite{Skovhus2026a}. For collinear magnets absent of spin-orbit coupling, Eq. \eqref{eq:collective mode diagonalization} decouples each collective mode from all other degrees of freedom, meaning that the generalized equation of motion \eqref{eq:generalized equation of motion} only involves the algebraic inverse of $\chi_n^{+-}(\mathbf{q},\omega) \equiv \langle v_n(\mathbf{q})| \chi^{+-}(\mathbf{q},\omega) |v_n(\mathbf{q})\rangle$ for each mode index $n$ and wave vector $\mathbf{q}$. Now, if $\chi_n^{+-}(\mathbf{q},\omega)$ is dominated by a \textit{single} magnon resonance at a complex frequency $z_n(\mathbf{q})$,
\begin{equation}
    \chi_n^{+-}(\mathbf{q},\omega) \simeq \frac{1}{\omega-z_n(\mathbf{q})} \lim_{\omega\rightarrow 0}\left[\omega-z_n(\mathbf{q})\right]\chi_n^{+-}(\mathbf{q},\omega),
    \label{eq:single magnon approximation}
\end{equation}
the Taylor expansion of its inverse can be truncated at linear order, and Eq. \eqref{eq:transverse qian-vignale} reduces to
\begin{equation}
    \left[\varepsilon_{\alpha\beta}\Omega_n(\mathbf{q}) + \delta_{\alpha\beta}\gamma_n(\mathbf{q})\right]\dot{m}^\beta_n(\mathbf{q},t) = \chi_n^{-1}(\mathbf{q})m^\alpha_n(\mathbf{q},t),
    \label{eq:magnon equation of motion}
\end{equation}
where $m^\alpha_n(\mathbf{q},t)=\int d\mathbf{r}\, e^{-i\mathbf{q}\cdot\mathbf{r}} v_{n\mathbf{q}}^*(\mathbf{r}) m^\alpha(\mathbf{r},t)$, $\chi_n^{-1}(\mathbf{q})$ denotes the $\omega\rightarrow0$ limit of $\chi_n^{-1}(\mathbf{q},\omega)=1/[2\chi_n^{+-}(\mathbf{q},\omega)]$ and
\begin{equation}
    \Omega_n(\mathbf{q}) = \mathrm{Re}\,\tilde{\Omega}_n(\mathbf{q}),
    \quad
    \gamma_n(\mathbf{q}) = \mathrm{Im}\,\tilde{\Omega}_n(\mathbf{q}),
\end{equation}
with
\begin{equation}
    \tilde{\Omega}_n(\mathbf{q}) = \lim_{\omega\rightarrow0}\frac{\partial}{\partial\omega}\frac{1}{2\chi_n^{+-}(\mathbf{q},\omega)}.
    \label{eq:differentiation omega}
\end{equation}
%
%

Not only is the magnon equation of motion \eqref{eq:magnon equation of motion} much more feasible to solve than Eq. \eqref{eq:transverse qian-vignale}, it also allows for straightforward interpretation of each term according to the quasi-particle picture. Rewriting Eq. \eqref{eq:magnon equation of motion}, the transverse magnetization is seen to follow the Newtonian dynamics of a damped harmonic oscillator,
\begin{align}
    \big[\Omega_n&(\mathbf{q})^2 + \gamma_n(\mathbf{q})^2\big]\ddot{m}_n^\alpha(\mathbf{q},t)
    \nonumber \\
    &=-\left[\chi_n^{-1}(\mathbf{q})\right]^2 m_n^\alpha(\mathbf{q},t)+2\chi_n^{-1}(\mathbf{q})\gamma_n(\mathbf{q})\dot{m}_n^\alpha(\mathbf{q},t),
    \label{eq:newtonian interpretation}
\end{align}
where $|\tilde{\Omega}_n(\mathbf{q})|^2$ represents the "mass" of the quasi-particle, which is subject to a restoring force with "spring constant" $[\chi_n^{-1}(\mathbf{q})]^2$ and friction with "damping coefficient" $-2\chi_n^{-1}(\mathbf{q})\gamma_n(\mathbf{q})$. 
While $\chi_n^{-1}(\mathbf{q})$ encodes the exchange interaction (see also Refs. \cite{Bruno2003,Antropov2003}), $\tilde{\Omega}_n(\mathbf{q})$ determines the amount of angular momentum initially transferred to the spin wave,
\begin{equation}
    m_n^\alpha(\mathbf{q},t=0^+)=\frac{\varepsilon_{\alpha\beta}\Omega_n(\mathbf{q})-\delta_{\alpha\beta}\gamma_n(\mathbf{q})}{|\tilde{\Omega}_n(\mathbf{q})|^2}W_{\mathrm{eff},n}^\beta(\mathbf{q},\omega=0).
    \label{eq:initial magnon conditions}
\end{equation}
Thus, it requires a greater external field strength to tilt the magnetization of a "heavier" magnon quasi-particle. 
Solving Eq. \eqref{eq:newtonian interpretation} subject to the single magnon approximation \eqref{eq:single magnon approximation}, one finds that $m_n^\alpha(\mathbf{q},t)=m_n^\perp(\mathbf{q}) e^{\mathrm{Im}\,z_n(\mathbf{q})t}\cos\left[\mathrm{Re}\,z_n(\mathbf{q})t -\phi_n^\alpha(\mathbf{q})\right]\theta(t)$, with $\phi_n^\alpha(\mathbf{q})$ and $m_n^\perp(\mathbf{q})$ determined by the initial conditions \eqref{eq:initial magnon conditions}. The assumptions \eqref{eq:collective mode diagonalization} and \eqref{eq:single magnon approximation} are thus consistent with a quasi-particle picture where $\mathrm{Re}\,z_n(\mathbf{q})$ represents the spin wave frequency, $-\mathrm{Im}\,z_n(\mathbf{q})$ its inverse lifetime, while 
$v_{n\mathbf{q}}(\mathbf{r})=v_{n\mathbf{q}}(\mathbf{r}+\mathbf{R})$ yields its sublattice representation. If the collective subspace is $\mathbf{q}$-independent, 
\begin{equation}
    \mathbb{I}_\mathrm{col}(\mathbf{q})=\sum_{n=0}^{N-1}|v_n(\mathbf{q})\rangle\langle v_n(\mathbf{q})| \simeq \mathbb{I}_\mathrm{col},
\end{equation}
the magnon equation of motion \eqref{eq:magnon equation of motion} uniquely defines from first principles the atomistic spin dynamics of the system. With $N$ magnetic site geometries $|a\rangle$ per unit cell chosen such as to span the collective magnon subspace, $|v_n(\mathbf{q})\rangle=\sum_a c_n^a(\mathbf{q})|a\rangle$, Eq. \eqref{eq:magnon equation of motion} represents the diagonal form of the atomistic equation of motion. 

In a practical sense, the magnon equation of motion \eqref{eq:magnon equation of motion} can be used to various different ends. One is to calculate the magnon frequency and lifetime dispersion from first principles. By approximating $\chi_n^{-1}(\mathbf{q})$ and $\tilde{\Omega}_n(\mathbf{q})$ at some level of band theory, one can directly estimate the complex quasi-particle frequency, 
%
\begin{equation}
       z_n(\mathbf{q}) \simeq -\frac{\chi_n^{-1}(\mathbf{q})}{\tilde{\Omega}_n(\mathbf{q})}.
       \label{eq:meom complex frequency solution}
\end{equation}
Another is to downfold a calculated or measured susceptibility $\chi^{+-}(\mathbf{q},\omega)$ to the quasi-particle picture of Eqs. \eqref{eq:collective mode diagonalization} and \eqref{eq:single magnon approximation}, 
and in turn estimate various derived physical properties. 
In this regard, a few comments are in order. 
As a retarded correlation function, the dynamic susceptibility is analytic in the upper-half complex frequency plane, where it can be expressed formally as a sum over energy eigenstates $\hat{H}_0|i\rangle=E_i|i\rangle$. Projected onto 
the collective subspace, 
%
\begin{equation}
    \chi_n^{+-}(\mathbf{q},\omega) = \lim_{\eta\rightarrow 0^+}\frac{1}{N_q}\sum_{i,j}\frac{(n_i-n_j)|\rho_{ij}^n(\mathbf{q})|^2}{\hbar\omega -(E_j-E_i) + i\hbar\eta},
    \label{eq:mode projected susceptibility}
\end{equation}
with state occupations $n_i$ at thermal equilibrium and matrix elements $\rho_{ij}^n(\mathbf{q})=\int d\mathbf{r}\,e^{-i\mathbf{q}\cdot\mathbf{r}}v_{n\mathbf{q}}^*(\mathbf{r}) \langle i|\hat{\psi}_\uparrow^\dagger(\mathbf{r})\hat{\psi}_\downarrow(\mathbf{r})|j\rangle$. In the zero temperature limit, the spectral weight of $\chi_n^{+-}(\mathbf{q},\omega)$ is distributed in two distinct ways; as isolated poles and in continuous segments spanning the bounds of the Stoner continuum. The isolated poles are undamped magnon resonances and occur at frequencies where the inverse susceptibility $\chi_n^{-1}(\mathbf{q},\omega)$ has a simple root. As a magnon pole enters the Stoner continuum, it is broadened into a peak of finite spectral width, and the corresponding root in the inverse susceptibility is displaced to the lower-half complex frequency plane. However, the root frequency $z_\mathrm{qp}$ is still rigorously defined via analytical continuation  
and may therefore be taken as the \textit{definition} of a quasi-particle resonance. 
The single magnon approximation \eqref{eq:single magnon approximation} assumes the spectrum of each magnon mode to be dominated by a single such resonance, that is, a single root in the inverse susceptibility,
\begin{equation}
    \chi_n^{-1}(\mathbf{q},\omega) \simeq \left[\omega-z_n(\mathbf{q})\right] \tilde{\Omega}^\mathrm{SMA}_n(\mathbf{q}).
    \label{eq:single root approximation}
\end{equation}
In order to guarantee that $z_n(\mathbf{q})$ constitutes an actual magnon resonance and that the correct spin stiffness is recovered in the static limit, we suggest to determine $\tilde{\Omega}^\mathrm{SMA}_n(\mathbf{q})$ by iteratively solving Eq. \eqref{eq:meom complex frequency solution} with
\begin{equation}
    \tilde{\Omega}^\mathrm{SMA}_n(\mathbf{q}) = \frac{\chi_n^{-1}(\mathbf{q},z_n(\mathbf{q}))-\chi_n^{-1}(\mathbf{q})}{z_n(\mathbf{q})},
    \label{eq:linear interpolation omega}
\end{equation}
such that $\chi_n^{-1}(\mathbf{q},z_n(\mathbf{q}))=0$ at self-consistency. In the long wavelength limit of the Goldstone mode, where $\lim_{q\rightarrow 0^+} z_0(q\hat{\mathbf{q}})=0$, Eq. \eqref{eq:linear interpolation omega} reduces to the first order frequency derivative \eqref{eq:differentiation omega}, but in general $\tilde{\Omega}_n(\mathbf{q})$ only yields an estimate of the self-consistent $\tilde{\Omega}^\mathrm{SMA}_n(\mathbf{q})$.


\begin{figure*}[tb]
    \centering
    \includegraphics[scale=1.0]{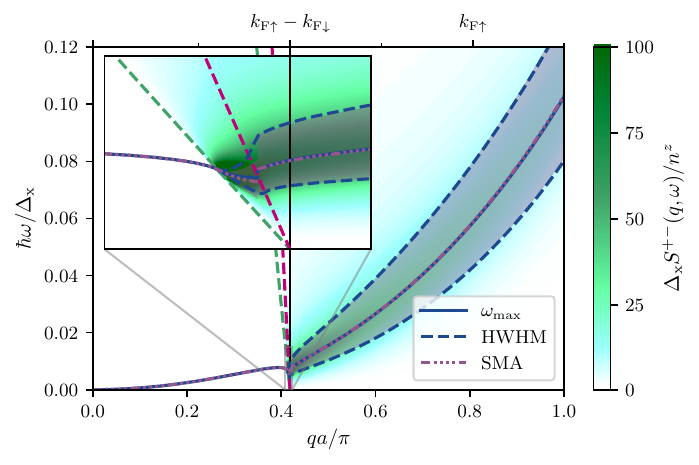}
    \caption{Scattering function $S^{+-}(q,\omega)=-\mathrm{Im}\,\chi^{+-}(q,\omega)/\pi$ of the Goldstone mode in a prototypical itinerant ferromagnet (colored contour). The global maximum (full blue line) and half-maxima (dashed blue lines) are compared to their respective counterparts (purple dash-dotted line and colored region) in the single magnon approximation \eqref{eq:single root approximation} constructed by solving Eqs. \eqref{eq:meom complex frequency solution} and \eqref{eq:linear interpolation omega} self-consistently. The dashed green and magenta lines indicate the derivative discontinuities at the Stoner bounds.
    }
    \label{fig:ifm_magnon_spectrum}
\end{figure*}
%
In Fig. \ref{fig:ifm_magnon_spectrum}, we present the magnon spectrum of the Goldstone mode in a prototypical itinerant ferromagnet and compare its spectral maximum and half-maxima to the downfolded spectrum in the single magnon approximation (SMA).
The model susceptibility is constructed by adding a localized spin contribution to the noninteracting susceptibility of the spin-polarized homogeneous electron gas within an RPA/ALDA treatment. For additional details on the model and its analytic continuation, please refer to the End Matter. 
Crucially, the model captures the main qualitative aspects of an itinerant ferromagnet: the quadratic Goldstone magnon dispersion in the long wavelength ($q\rightarrow0$) limit and the broadening of the magnon peak upon entry into the Stoner continuum (Landau damping). 
Evidently, the SMA retains a virtually perfect description of both. For wave vectors $q$, where the magnon frequency resides below the bounds of the Stoner continuum, the pole position of the undamped magnon is reproduced exactly, 
and even in the Landau damped regime inside the Stoner continuum, both peak position and width are extremely well reproduced.
This simply means that the susceptibility \textit{is} dominated by a single magnon resonance (somewhat by construction), and, as a result, that the dynamics follow the magnon equation of motion \eqref{eq:magnon equation of motion}.

\begin{figure*}[tb]
    \centering
    \includegraphics[scale=1.0]{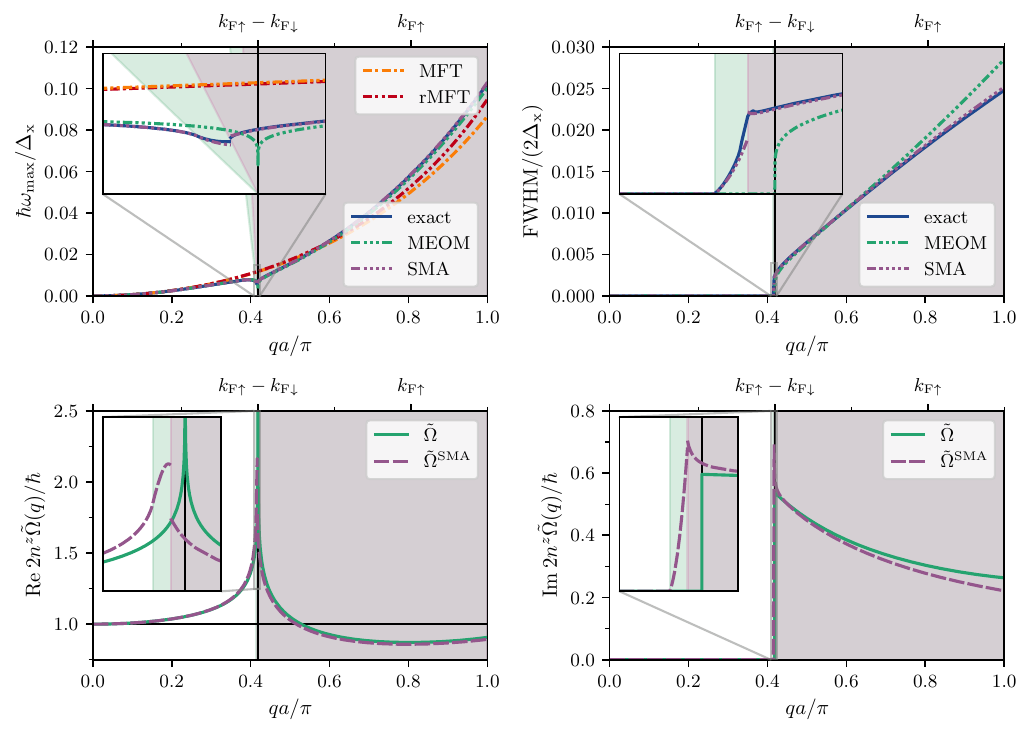}
    \caption{Top row: performance of the bare MEOM dispersion \eqref{eq:meom complex frequency solution} and the downfolded SMA \eqref{eq:single root approximation} in comparison with Heisenberg model approaches regarding the global maximum (left) and the full width at half maximum (right) of the scattering function. Bottom row: comparison of the real (left) and imaginary (right) parts of $\tilde{\Omega}$, evaluated as a derivative in the static limit \eqref{eq:differentiation omega} and self-consistently according to the SMA \eqref{eq:linear interpolation omega}. The colored regions illustrate where the magnon resonance is situated with respect to the Stoner bounds.}
    \label{fig:ifm_magnon_equation_of_motion}
\end{figure*}
By virtue of its single magnon resonance, the present model also allows us to further discuss the physics of each of the three coefficients $\chi^{-1}(\mathbf{q})$, $\Omega(\mathbf{q})$ and $\gamma(\mathbf{q})$, and to compare estimates of the magnon resonance frequencies through Eq. \eqref{eq:meom complex frequency solution} to more traditional methods based on a mapping to the Heisenberg model. In order for such a mapping to be exact, the electron-magnon coupling needs to vanish. This can be realized in the insulating limit of the present model, where the Stoner excitations are completely quenched from the scattering function due to the narrow spectral distribution of the localized $d$ or $f$ electronic states \cite{Skovhus2026a}. In this limit, the magnon resonance carries all of the fixed spectral weight $n^z$, 
\begin{equation}
     \chi^{+-}(\mathbf{q},\omega) \simeq \lim_{\eta\rightarrow 0^+} \frac{n^z}{\hbar\omega-\hbar\omega(\mathbf{q})+i\hbar\eta},
\end{equation}
and the quasi-particle mass becomes nondispersive, $\tilde{\Omega}(\mathbf{q})=\tilde{\Omega}(\mathbf{0})=\hbar/(2n^z)$. The only nontrivial coefficient left is the Heisenberg exchange encoded in $\chi^{-1}(\mathbf{q})$, and the magnon equation of motion \eqref{eq:magnon equation of motion} reduces to the linearized Landau-Lifshitz equation (without damping). If effects of the Stoner continuum are (re)introduced, $\tilde{\Omega}(\mathbf{q})$ will start to deviate from $\tilde{\Omega}(\mathbf{0})$, and the Heisenberg model description will break down due to the electron-magnon coupling that this implies.  
%
In Fig. \ref{fig:ifm_magnon_equation_of_motion}, we compare the actual peak dispersion of our prototypical itinerant ferromagnet to the downfolded SMA, the bare magnon equation of motion (MEOM), the Heisenberg model dispersion resulting from application of the magnetic force theorem (MFT) according to Ref. \cite{Liechtenstein1987}, $\hbar\omega_\mathrm{MFT}(\mathbf{q})= -\Delta_\mathrm{x}\left[\chi_\mathrm{KS}^{+-}(\mathbf{0}) - \chi_\mathrm{KS}^{+-}(\mathbf{q})\right]\Delta_\mathrm{x}/n^z$ \cite{Durhuus2023,Skovhus2025}, and the insulating limit dispersion $\hbar\omega(\mathbf{q})=-2n^z \chi^{-1}(\mathbf{q})$ which is identical to the renormalized (rMFT) dispersion of Ref. \cite{Bruno2003}. 
In the long wavelength limit, the electron-magnon coupling $\tilde{\Omega}(\mathbf{q})-\tilde{\Omega}(\mathbf{0})$ vanishes, and all methods reproduce the exact curvature at $q=0$ (the spin wave stiffness). 
However, whereas $\chi^{-1}(\mathbf{q})$ and $\chi_\mathrm{KS}^{+-}(\mathbf{q})$ both are smooth and monotonic as a function of $q$ (the exchange interaction is well localized in real space), the actual magnon dispersion is neither due to abrupt changes in the electron-magnon coupling (which is long ranged in real space) around the magnon's entry into the Stoner continuum. Here $\Omega(\mathbf{q})$ is sharply peaked, resulting in a momentary redshift of the dispersion, see Fig. \ref{fig:ifm_magnon_equation_of_motion} and Eq. \eqref{eq:meom complex frequency solution}, and at the entry itself $\gamma(\mathbf{q})$ becomes finite, resulting in a finite magnon lifetime (Landau damping). Neither of these effects are captured at the Heisenberg model level, but well described using the magnon equation of motion \eqref{eq:magnon equation of motion}, especially in the self-consistent SMA. The primary effects of self-consistency are a shift in the Landau damping onset from $q=k_{\mathrm{F}\uparrow}-k_{\mathrm{F}\downarrow}$ (the minimal $q$ which connects the majority and minority Fermi surfaces) to the point where the magnon resonance actually enters the Stoner continuum (at finite $\omega$), along with an improvement of the inferred lifetime at large $q$.


Assuming that magnons belong to distinct collective modes of the dynamic susceptibility, with one magnon resonance per mode, we have derived the magnon equation of motion. In addition to Heisenberg exchange, the equation of motion is subject to electron-magnon coupling via the Stoner continuum of single-particle excitations. The real part of the coupling (the dispersive "Berry curvature") renormalizes the magnon dispersion, while the imaginary part is responsible for Landau damping. By including both effects of the electron-magnon coupling, the magnon spectrum of prototypical itinerant ferromagnets can be described accurately and self-consistently. This is a necessity, if one wants control over damping effects in magnon spintronics \cite{Wolf2001,Neusser2009,Lenk2011,Chumak2015,Qin2015,Schoen2016,Paischer2024}, to accurate model magnetic phase transitions \cite{Paischer2021,RajeevPavizhakumari2025} and ultrafast demagnetization effects \cite{Beaurepaire1996,Koopmans2005,Kirilyuk2010,Pankratova2022}, or to self-consistently account for electron-magnon coupling in calculations of the electronic band structure \cite{Nabok2021,Paischer2023}.

\vspace{10pt}
\textit{Acknowledgments}---The work presented here is supported by the Carlsberg Foundation, grant CF24-1413.

\vspace{10pt}
\textit{Data availability}---The data that support the findings of this paper are publicly available \cite{meomdata}.

\bibliography{bibliography}

\begin{thebibliography}{49}%
\makeatletter
\providecommand \@ifxundefined [1]{%
 \@ifx{#1\undefined}
}%
\providecommand \@ifnum [1]{%
 \ifnum #1\expandafter \@firstoftwo
 \else \expandafter \@secondoftwo
 \fi
}%
\providecommand \@ifx [1]{%
 \ifx #1\expandafter \@firstoftwo
 \else \expandafter \@secondoftwo
 \fi
}%
\providecommand \natexlab [1]{#1}%
\providecommand \enquote  [1]{``#1''}%
\providecommand \bibnamefont  [1]{#1}%
\providecommand \bibfnamefont [1]{#1}%
\providecommand \citenamefont [1]{#1}%
\providecommand \href@noop [0]{\@secondoftwo}%
\providecommand \href [0]{\begingroup \@sanitize@url \@href}%
\providecommand \@href[1]{\@@startlink{#1}\@@href}%
\providecommand \@@href[1]{\endgroup#1\@@endlink}%
\providecommand \@sanitize@url [0]{\catcode `\\12\catcode `\$12\catcode
  `\&12\catcode `\#12\catcode `\^12\catcode `\_12\catcode `\%12\relax}%
\providecommand \@@startlink[1]{}%
\providecommand \@@endlink[0]{}%
\providecommand \url  [0]{\begingroup\@sanitize@url \@url }%
\providecommand \@url [1]{\endgroup\@href {#1}{\urlprefix }}%
\providecommand \urlprefix  [0]{URL }%
\providecommand \Eprint [0]{\href }%
\providecommand \doibase [0]{https://doi.org/}%
\providecommand \selectlanguage [0]{\@gobble}%
\providecommand \bibinfo  [0]{\@secondoftwo}%
\providecommand \bibfield  [0]{\@secondoftwo}%
\providecommand \translation [1]{[#1]}%
\providecommand \BibitemOpen [0]{}%
\providecommand \bibitemStop [0]{}%
\providecommand \bibitemNoStop [0]{.\EOS\space}%
\providecommand \EOS [0]{\spacefactor3000\relax}%
\providecommand \BibitemShut  [1]{\csname bibitem#1\endcsname}%
\let\auto@bib@innerbib\@empty
\bibitem [{\citenamefont {Liechtenstein}\ \emph {et~al.}(1987)\citenamefont
  {Liechtenstein}, \citenamefont {Katsnelson}, \citenamefont {Antropov},\ and\
  \citenamefont {Gubanov}}]{Liechtenstein1987}%
  \BibitemOpen
  \bibfield  {author} {\bibinfo {author} {\bibfnamefont {A.}~\bibnamefont
  {Liechtenstein}}, \bibinfo {author} {\bibfnamefont {M.}~\bibnamefont
  {Katsnelson}}, \bibinfo {author} {\bibfnamefont {V.}~\bibnamefont
  {Antropov}},\ and\ \bibinfo {author} {\bibfnamefont {V.}~\bibnamefont
  {Gubanov}},\ }\href {https://doi.org/10.1016/0304-8853(87)90721-9} {\bibfield
   {journal} {\bibinfo  {journal} {J. Magn. Magn. Mater.}\ }\textbf {\bibinfo
  {volume} {67}},\ \bibinfo {pages} {65} (\bibinfo {year} {1987})}\BibitemShut
  {NoStop}%
\bibitem [{\citenamefont {Halilov}\ \emph {et~al.}(1998)\citenamefont
  {Halilov}, \citenamefont {Eschrig}, \citenamefont {Perlov},\ and\
  \citenamefont {Oppeneer}}]{Halilov1998}%
  \BibitemOpen
  \bibfield  {author} {\bibinfo {author} {\bibfnamefont {S.~V.}\ \bibnamefont
  {Halilov}}, \bibinfo {author} {\bibfnamefont {H.}~\bibnamefont {Eschrig}},
  \bibinfo {author} {\bibfnamefont {A.~Y.}\ \bibnamefont {Perlov}},\ and\
  \bibinfo {author} {\bibfnamefont {P.~M.}\ \bibnamefont {Oppeneer}},\ }\href
  {https://doi.org/10.1103/PhysRevB.58.293} {\bibfield  {journal} {\bibinfo
  {journal} {Physical Review B}\ }\textbf {\bibinfo {volume} {58}},\ \bibinfo
  {pages} {293} (\bibinfo {year} {1998})}\BibitemShut {NoStop}%
\bibitem [{\citenamefont {Grotheer}\ \emph {et~al.}(2001)\citenamefont
  {Grotheer}, \citenamefont {Ederer},\ and\ \citenamefont
  {F\"{a}hnle}}]{Grotheer2001}%
  \BibitemOpen
  \bibfield  {author} {\bibinfo {author} {\bibfnamefont {O.}~\bibnamefont
  {Grotheer}}, \bibinfo {author} {\bibfnamefont {C.}~\bibnamefont {Ederer}},\
  and\ \bibinfo {author} {\bibfnamefont {M.}~\bibnamefont {F\"{a}hnle}},\
  }\href {https://doi.org/10.1103/PhysRevB.63.100401} {\bibfield  {journal}
  {\bibinfo  {journal} {Physical Review B}\ }\textbf {\bibinfo {volume} {63}},\
  \bibinfo {pages} {100401} (\bibinfo {year} {2001})}\BibitemShut {NoStop}%
\bibitem [{\citenamefont {Antropov}(2003)}]{Antropov2003}%
  \BibitemOpen
  \bibfield  {author} {\bibinfo {author} {\bibfnamefont {V.}~\bibnamefont
  {Antropov}},\ }\href {https://doi.org/10.1016/S0304-8853(03)00206-3}
  {\bibfield  {journal} {\bibinfo  {journal} {Journal of Magnetism and Magnetic
  Materials}\ }\textbf {\bibinfo {volume} {262}},\ \bibinfo {pages} {L192}
  (\bibinfo {year} {2003})}\BibitemShut {NoStop}%
\bibitem [{\citenamefont {Bruno}(2003)}]{Bruno2003}%
  \BibitemOpen
  \bibfield  {author} {\bibinfo {author} {\bibfnamefont {P.}~\bibnamefont
  {Bruno}},\ }\href {https://doi.org/10.1103/PhysRevLett.90.087205} {\bibfield
  {journal} {\bibinfo  {journal} {Physical Review Letters}\ }\textbf {\bibinfo
  {volume} {90}},\ \bibinfo {pages} {087205} (\bibinfo {year}
  {2003})}\BibitemShut {NoStop}%
\bibitem [{\citenamefont {Szilva}\ \emph {et~al.}(2023)\citenamefont {Szilva},
  \citenamefont {Kvashnin}, \citenamefont {Stepanov}, \citenamefont
  {Nordstr\"{o}m}, \citenamefont {Eriksson}, \citenamefont {Lichtenstein},\
  and\ \citenamefont {Katsnelson}}]{Szilva2023}%
  \BibitemOpen
  \bibfield  {author} {\bibinfo {author} {\bibfnamefont {A.}~\bibnamefont
  {Szilva}}, \bibinfo {author} {\bibfnamefont {Y.}~\bibnamefont {Kvashnin}},
  \bibinfo {author} {\bibfnamefont {E.~A.}\ \bibnamefont {Stepanov}}, \bibinfo
  {author} {\bibfnamefont {L.}~\bibnamefont {Nordstr\"{o}m}}, \bibinfo {author}
  {\bibfnamefont {O.}~\bibnamefont {Eriksson}}, \bibinfo {author}
  {\bibfnamefont {A.~I.}\ \bibnamefont {Lichtenstein}},\ and\ \bibinfo {author}
  {\bibfnamefont {M.~I.}\ \bibnamefont {Katsnelson}},\ }\href
  {https://doi.org/10.1103/RevModPhys.95.035004} {\bibfield  {journal}
  {\bibinfo  {journal} {Reviews of Modern Physics}\ }\textbf {\bibinfo {volume}
  {95}},\ \bibinfo {pages} {035004} (\bibinfo {year} {2023})}\BibitemShut
  {NoStop}%
\bibitem [{\citenamefont {Solovyev}(2024)}]{Solovyev2024}%
  \BibitemOpen
  \bibfield  {author} {\bibinfo {author} {\bibfnamefont {I.~V.}\ \bibnamefont
  {Solovyev}},\ }\href {https://doi.org/10.1088/1361-648X/ad215a} {\bibfield
  {journal} {\bibinfo  {journal} {Journal of Physics: Condensed Matter}\
  }\textbf {\bibinfo {volume} {36}},\ \bibinfo {pages} {223001} (\bibinfo
  {year} {2024})}\BibitemShut {NoStop}%
\bibitem [{\citenamefont {Moriya}(1985)}]{Moriya1985}%
  \BibitemOpen
  \bibfield  {author} {\bibinfo {author} {\bibfnamefont {T.}~\bibnamefont
  {Moriya}},\ }\href {https://doi.org/10.1007/978-3-642-82499-9} {\emph
  {\bibinfo {title} {Spin Fluctuations in Itinerant Electron Magnetism}}},\
  \bibinfo {series} {Springer Series in Solid-State Sciences}, Vol.~\bibinfo
  {volume} {56}\ (\bibinfo  {publisher} {Springer-Verlag Berlin Heidelberg},\
  \bibinfo {year} {1985})\BibitemShut {NoStop}%
\bibitem [{\citenamefont {Niu}\ and\ \citenamefont {Kleinman}(1998)}]{Niu1998}%
  \BibitemOpen
  \bibfield  {author} {\bibinfo {author} {\bibfnamefont {Q.}~\bibnamefont
  {Niu}}\ and\ \bibinfo {author} {\bibfnamefont {L.}~\bibnamefont {Kleinman}},\
  }\href {https://doi.org/10.1103/PhysRevLett.80.2205} {\bibfield  {journal}
  {\bibinfo  {journal} {Physical Review Letters}\ }\textbf {\bibinfo {volume}
  {80}},\ \bibinfo {pages} {2205} (\bibinfo {year} {1998})}\BibitemShut
  {NoStop}%
\bibitem [{\citenamefont {Niu}\ \emph {et~al.}(1999)\citenamefont {Niu},
  \citenamefont {Wang}, \citenamefont {Kleinman}, \citenamefont {Liu},
  \citenamefont {Nicholson},\ and\ \citenamefont {Stocks}}]{Niu1999}%
  \BibitemOpen
  \bibfield  {author} {\bibinfo {author} {\bibfnamefont {Q.}~\bibnamefont
  {Niu}}, \bibinfo {author} {\bibfnamefont {X.}~\bibnamefont {Wang}}, \bibinfo
  {author} {\bibfnamefont {L.}~\bibnamefont {Kleinman}}, \bibinfo {author}
  {\bibfnamefont {W.-M.}\ \bibnamefont {Liu}}, \bibinfo {author} {\bibfnamefont
  {D.~M.~C.}\ \bibnamefont {Nicholson}},\ and\ \bibinfo {author} {\bibfnamefont
  {G.~M.}\ \bibnamefont {Stocks}},\ }\href
  {https://doi.org/10.1103/PhysRevLett.83.207} {\bibfield  {journal} {\bibinfo
  {journal} {Physical Review Letters}\ }\textbf {\bibinfo {volume} {83}},\
  \bibinfo {pages} {207} (\bibinfo {year} {1999})}\BibitemShut {NoStop}%
\bibitem [{\citenamefont {Savrasov}(1998)}]{Savrasov1998}%
  \BibitemOpen
  \bibfield  {author} {\bibinfo {author} {\bibfnamefont {S.~Y.}\ \bibnamefont
  {Savrasov}},\ }\href {https://doi.org/10.1103/PhysRevLett.81.2570} {\bibfield
   {journal} {\bibinfo  {journal} {Physical Review Letters}\ }\textbf {\bibinfo
  {volume} {81}},\ \bibinfo {pages} {2570} (\bibinfo {year}
  {1998})}\BibitemShut {NoStop}%
\bibitem [{\citenamefont {Aryasetiawan}\ and\ \citenamefont
  {Karlsson}(1999)}]{Aryasetiawan1999}%
  \BibitemOpen
  \bibfield  {author} {\bibinfo {author} {\bibfnamefont {F.}~\bibnamefont
  {Aryasetiawan}}\ and\ \bibinfo {author} {\bibfnamefont {K.}~\bibnamefont
  {Karlsson}},\ }\href {https://doi.org/10.1103/PhysRevB.60.7419} {\bibfield
  {journal} {\bibinfo  {journal} {Physical Review B}\ }\textbf {\bibinfo
  {volume} {60}},\ \bibinfo {pages} {7419} (\bibinfo {year}
  {1999})}\BibitemShut {NoStop}%
\bibitem [{\citenamefont {Karlsson}\ and\ \citenamefont
  {Aryasetiawan}(2000)}]{Karlsson2000}%
  \BibitemOpen
  \bibfield  {author} {\bibinfo {author} {\bibfnamefont {K.}~\bibnamefont
  {Karlsson}}\ and\ \bibinfo {author} {\bibfnamefont {F.}~\bibnamefont
  {Aryasetiawan}},\ }\href {https://doi.org/10.1103/PhysRevB.62.3006}
  {\bibfield  {journal} {\bibinfo  {journal} {Physical Review B}\ }\textbf
  {\bibinfo {volume} {62}},\ \bibinfo {pages} {3006} (\bibinfo {year}
  {2000})}\BibitemShut {NoStop}%
\bibitem [{\citenamefont {\ifmmode \mbox{\c{S}}\else \c{S}\fi{}a\ifmmode
  \mbox{\c{s}}\else \c{s}\fi{}\ifmmode \imath \else \i
  \fi{}o\ifmmode~\breve{g}\else \u{g}\fi{}lu}\ \emph
  {et~al.}(2010)\citenamefont {\ifmmode \mbox{\c{S}}\else \c{S}\fi{}a\ifmmode
  \mbox{\c{s}}\else \c{s}\fi{}\ifmmode \imath \else \i
  \fi{}o\ifmmode~\breve{g}\else \u{g}\fi{}lu}, \citenamefont {Schindlmayr},
  \citenamefont {Friedrich}, \citenamefont {Freimuth},\ and\ \citenamefont
  {Bl\"ugel}}]{SasIoglu2010}%
  \BibitemOpen
  \bibfield  {author} {\bibinfo {author} {\bibfnamefont {E.}~\bibnamefont
  {\ifmmode \mbox{\c{S}}\else \c{S}\fi{}a\ifmmode \mbox{\c{s}}\else
  \c{s}\fi{}\ifmmode \imath \else \i \fi{}o\ifmmode~\breve{g}\else
  \u{g}\fi{}lu}}, \bibinfo {author} {\bibfnamefont {A.}~\bibnamefont
  {Schindlmayr}}, \bibinfo {author} {\bibfnamefont {C.}~\bibnamefont
  {Friedrich}}, \bibinfo {author} {\bibfnamefont {F.}~\bibnamefont
  {Freimuth}},\ and\ \bibinfo {author} {\bibfnamefont {S.}~\bibnamefont
  {Bl\"ugel}},\ }\href {https://doi.org/10.1103/PhysRevB.81.054434} {\bibfield
  {journal} {\bibinfo  {journal} {Phys. Rev. B}\ }\textbf {\bibinfo {volume}
  {81}},\ \bibinfo {pages} {054434} (\bibinfo {year} {2010})}\BibitemShut
  {NoStop}%
\bibitem [{\citenamefont {Buczek}\ \emph {et~al.}(2011)\citenamefont {Buczek},
  \citenamefont {Ernst},\ and\ \citenamefont {Sandratskii}}]{Buczek2011b}%
  \BibitemOpen
  \bibfield  {author} {\bibinfo {author} {\bibfnamefont {P.}~\bibnamefont
  {Buczek}}, \bibinfo {author} {\bibfnamefont {A.}~\bibnamefont {Ernst}},\ and\
  \bibinfo {author} {\bibfnamefont {L.~M.}\ \bibnamefont {Sandratskii}},\
  }\href {https://doi.org/10.1103/PhysRevB.84.174418} {\bibfield  {journal}
  {\bibinfo  {journal} {Physical Review B}\ }\textbf {\bibinfo {volume} {84}},\
  \bibinfo {pages} {174418} (\bibinfo {year} {2011})}\BibitemShut {NoStop}%
\bibitem [{\citenamefont {Lounis}\ \emph {et~al.}(2011)\citenamefont {Lounis},
  \citenamefont {Costa}, \citenamefont {Muniz},\ and\ \citenamefont
  {Mills}}]{Lounis2011}%
  \BibitemOpen
  \bibfield  {author} {\bibinfo {author} {\bibfnamefont {S.}~\bibnamefont
  {Lounis}}, \bibinfo {author} {\bibfnamefont {A.~T.}\ \bibnamefont {Costa}},
  \bibinfo {author} {\bibfnamefont {R.~B.}\ \bibnamefont {Muniz}},\ and\
  \bibinfo {author} {\bibfnamefont {D.~L.}\ \bibnamefont {Mills}},\ }\href
  {https://doi.org/10.1103/PhysRevB.83.035109} {\bibfield  {journal} {\bibinfo
  {journal} {Physical Review B}\ }\textbf {\bibinfo {volume} {83}},\ \bibinfo
  {pages} {035109} (\bibinfo {year} {2011})}\BibitemShut {NoStop}%
\bibitem [{\citenamefont {Rousseau}\ \emph {et~al.}(2012)\citenamefont
  {Rousseau}, \citenamefont {Eiguren},\ and\ \citenamefont
  {Bergara}}]{Rousseau2012}%
  \BibitemOpen
  \bibfield  {author} {\bibinfo {author} {\bibfnamefont {B.}~\bibnamefont
  {Rousseau}}, \bibinfo {author} {\bibfnamefont {A.}~\bibnamefont {Eiguren}},\
  and\ \bibinfo {author} {\bibfnamefont {A.}~\bibnamefont {Bergara}},\ }\href
  {https://doi.org/10.1103/PhysRevB.85.054305} {\bibfield  {journal} {\bibinfo
  {journal} {Physical Review B}\ }\textbf {\bibinfo {volume} {85}},\ \bibinfo
  {pages} {054305} (\bibinfo {year} {2012})}\BibitemShut {NoStop}%
\bibitem [{\citenamefont {M{\"{u}}ller}\ \emph {et~al.}(2016)\citenamefont
  {M{\"{u}}ller}, \citenamefont {Friedrich},\ and\ \citenamefont
  {Bl{\"{u}}gel}}]{Muller2016}%
  \BibitemOpen
  \bibfield  {author} {\bibinfo {author} {\bibfnamefont {M.~C. T.~D.}\
  \bibnamefont {M{\"{u}}ller}}, \bibinfo {author} {\bibfnamefont
  {C.}~\bibnamefont {Friedrich}},\ and\ \bibinfo {author} {\bibfnamefont
  {S.}~\bibnamefont {Bl{\"{u}}gel}},\ }\href
  {https://doi.org/10.1103/PhysRevB.94.064433} {\bibfield  {journal} {\bibinfo
  {journal} {Physical Review B}\ }\textbf {\bibinfo {volume} {94}},\ \bibinfo
  {pages} {064433} (\bibinfo {year} {2016})}\BibitemShut {NoStop}%
\bibitem [{\citenamefont {Cao}\ \emph {et~al.}(2018)\citenamefont {Cao},
  \citenamefont {Lambert}, \citenamefont {Radaelli},\ and\ \citenamefont
  {Giustino}}]{Cao2017}%
  \BibitemOpen
  \bibfield  {author} {\bibinfo {author} {\bibfnamefont {K.}~\bibnamefont
  {Cao}}, \bibinfo {author} {\bibfnamefont {H.}~\bibnamefont {Lambert}},
  \bibinfo {author} {\bibfnamefont {P.~G.}\ \bibnamefont {Radaelli}},\ and\
  \bibinfo {author} {\bibfnamefont {F.}~\bibnamefont {Giustino}},\ }\href
  {https://doi.org/10.1103/PhysRevB.97.024420} {\bibfield  {journal} {\bibinfo
  {journal} {Physical Review B}\ }\textbf {\bibinfo {volume} {97}},\ \bibinfo
  {pages} {024420} (\bibinfo {year} {2018})}\BibitemShut {NoStop}%
\bibitem [{\citenamefont {Singh}\ \emph {et~al.}(2019)\citenamefont {Singh},
  \citenamefont {Elliott}, \citenamefont {Nautiyal}, \citenamefont {Dewhurst},\
  and\ \citenamefont {Sharma}}]{Singh2019}%
  \BibitemOpen
  \bibfield  {author} {\bibinfo {author} {\bibfnamefont {N.}~\bibnamefont
  {Singh}}, \bibinfo {author} {\bibfnamefont {P.}~\bibnamefont {Elliott}},
  \bibinfo {author} {\bibfnamefont {T.}~\bibnamefont {Nautiyal}}, \bibinfo
  {author} {\bibfnamefont {J.~K.}\ \bibnamefont {Dewhurst}},\ and\ \bibinfo
  {author} {\bibfnamefont {S.}~\bibnamefont {Sharma}},\ }\href
  {https://doi.org/10.1103/PhysRevB.99.035151} {\bibfield  {journal} {\bibinfo
  {journal} {Physical Review B}\ }\textbf {\bibinfo {volume} {99}},\ \bibinfo
  {pages} {035151} (\bibinfo {year} {2019})}\BibitemShut {NoStop}%
\bibitem [{\citenamefont {Okumura}\ \emph {et~al.}(2019)\citenamefont
  {Okumura}, \citenamefont {Sato},\ and\ \citenamefont {Kotani}}]{Okumura2019}%
  \BibitemOpen
  \bibfield  {author} {\bibinfo {author} {\bibfnamefont {H.}~\bibnamefont
  {Okumura}}, \bibinfo {author} {\bibfnamefont {K.}~\bibnamefont {Sato}},\ and\
  \bibinfo {author} {\bibfnamefont {T.}~\bibnamefont {Kotani}},\ }\href
  {https://doi.org/10.1103/PhysRevB.100.054419} {\bibfield  {journal} {\bibinfo
   {journal} {Physical Review B}\ }\textbf {\bibinfo {volume} {100}},\ \bibinfo
  {pages} {054419} (\bibinfo {year} {2019})}\BibitemShut {NoStop}%
\bibitem [{\citenamefont {Tancogne-Dejean}\ \emph {et~al.}(2020)\citenamefont
  {Tancogne-Dejean}, \citenamefont {Eich},\ and\ \citenamefont
  {Rubio}}]{Tancogne-Dejean2020}%
  \BibitemOpen
  \bibfield  {author} {\bibinfo {author} {\bibfnamefont {N.}~\bibnamefont
  {Tancogne-Dejean}}, \bibinfo {author} {\bibfnamefont {F.~G.}\ \bibnamefont
  {Eich}},\ and\ \bibinfo {author} {\bibfnamefont {A.}~\bibnamefont {Rubio}},\
  }\href {https://doi.org/10.1021/acs.jctc.9b01064} {\bibfield  {journal}
  {\bibinfo  {journal} {Journal of Chemical Theory and Computation}\ }\textbf
  {\bibinfo {volume} {16}},\ \bibinfo {pages} {1007} (\bibinfo {year}
  {2020})}\BibitemShut {NoStop}%
\bibitem [{\citenamefont {Friedrich}\ \emph {et~al.}(2020)\citenamefont
  {Friedrich}, \citenamefont {M{\"{u}}ller},\ and\ \citenamefont
  {Bl{\"{u}}gel}}]{Friedrich2020}%
  \BibitemOpen
  \bibfield  {author} {\bibinfo {author} {\bibfnamefont {C.}~\bibnamefont
  {Friedrich}}, \bibinfo {author} {\bibfnamefont {M.~C. T.~D.}\ \bibnamefont
  {M{\"{u}}ller}},\ and\ \bibinfo {author} {\bibfnamefont {S.}~\bibnamefont
  {Bl{\"{u}}gel}},\ }in\ \href {https://doi.org/10.1007/978-3-319-44677-6_74}
  {\emph {\bibinfo {booktitle} {Handbook of Materials Modeling: Methods: Theory
  and Modeling}}},\ \bibinfo {editor} {edited by\ \bibinfo {editor}
  {\bibfnamefont {W.}~\bibnamefont {Andreoni}}\ and\ \bibinfo {editor}
  {\bibfnamefont {S.}~\bibnamefont {Yip}}}\ (\bibinfo  {publisher} {Springer
  International Publishing},\ \bibinfo {address} {Cham},\ \bibinfo {year}
  {2020})\ pp.\ \bibinfo {pages} {919--956}\BibitemShut {NoStop}%
\bibitem [{\citenamefont {Skovhus}\ and\ \citenamefont
  {Olsen}(2021)}]{Skovhus2021}%
  \BibitemOpen
  \bibfield  {author} {\bibinfo {author} {\bibfnamefont {T.}~\bibnamefont
  {Skovhus}}\ and\ \bibinfo {author} {\bibfnamefont {T.}~\bibnamefont
  {Olsen}},\ }\href {https://doi.org/10.1103/PhysRevB.103.245110} {\bibfield
  {journal} {\bibinfo  {journal} {Physical Review B}\ }\textbf {\bibinfo
  {volume} {103}},\ \bibinfo {pages} {245110} (\bibinfo {year}
  {2021})}\BibitemShut {NoStop}%
\bibitem [{\citenamefont {Liu}\ \emph {et~al.}(2023)\citenamefont {Liu},
  \citenamefont {Lin},\ and\ \citenamefont {Feng}}]{Liu2023}%
  \BibitemOpen
  \bibfield  {author} {\bibinfo {author} {\bibfnamefont {X.}~\bibnamefont
  {Liu}}, \bibinfo {author} {\bibfnamefont {Y.}~\bibnamefont {Lin}},\ and\
  \bibinfo {author} {\bibfnamefont {J.}~\bibnamefont {Feng}},\ }\href
  {https://doi.org/10.1103/PhysRevB.108.094405} {\bibfield  {journal} {\bibinfo
   {journal} {Physical Review B}\ }\textbf {\bibinfo {volume} {108}},\ \bibinfo
  {pages} {094405} (\bibinfo {year} {2023})}\BibitemShut {NoStop}%
\bibitem [{\citenamefont {Skovhus}\ and\ \citenamefont
  {Olsen}(2026)}]{Skovhus2026a}%
  \BibitemOpen
  \bibfield  {author} {\bibinfo {author} {\bibfnamefont {T.}~\bibnamefont
  {Skovhus}}\ and\ \bibinfo {author} {\bibfnamefont {T.}~\bibnamefont
  {Olsen}},\ }\href {https://arxiv.org/abs/2604.22484} {\bibinfo {title}
  {Classifying magnons in itinerant ferromagnets from linear response tddft:
  Fe, ni and co revisited}} (\bibinfo {year} {2026}),\ \Eprint
  {https://arxiv.org/abs/2604.22484} {arXiv:2604.22484 [cond-mat.mtrl-sci]}
  \BibitemShut {NoStop}%
\bibitem [{\citenamefont {Qian}\ and\ \citenamefont
  {Vignale}(2002)}]{Qian2002}%
  \BibitemOpen
  \bibfield  {author} {\bibinfo {author} {\bibfnamefont {Z.}~\bibnamefont
  {Qian}}\ and\ \bibinfo {author} {\bibfnamefont {G.}~\bibnamefont {Vignale}},\
  }\href {https://doi.org/10.1103/PhysRevLett.88.056404} {\bibfield  {journal}
  {\bibinfo  {journal} {Physical Review Letters}\ }\textbf {\bibinfo {volume}
  {88}},\ \bibinfo {pages} {056404} (\bibinfo {year} {2002})}\BibitemShut
  {NoStop}%
\bibitem [{\citenamefont {Skovhus}\ and\ \citenamefont
  {Olsen}(2022)}]{Skovhus2022}%
  \BibitemOpen
  \bibfield  {author} {\bibinfo {author} {\bibfnamefont {T.}~\bibnamefont
  {Skovhus}}\ and\ \bibinfo {author} {\bibfnamefont {T.}~\bibnamefont
  {Olsen}},\ }\href {https://doi.org/10.1103/PhysRevB.106.085131} {\bibfield
  {journal} {\bibinfo  {journal} {Physical Review B}\ }\textbf {\bibinfo
  {volume} {106}},\ \bibinfo {pages} {085131} (\bibinfo {year}
  {2022})}\BibitemShut {NoStop}%
\bibitem [{\citenamefont {Durhuus}\ \emph {et~al.}(2023)\citenamefont
  {Durhuus}, \citenamefont {Skovhus},\ and\ \citenamefont
  {Olsen}}]{Durhuus2023}%
  \BibitemOpen
  \bibfield  {author} {\bibinfo {author} {\bibfnamefont {F.~L.}\ \bibnamefont
  {Durhuus}}, \bibinfo {author} {\bibfnamefont {T.}~\bibnamefont {Skovhus}},\
  and\ \bibinfo {author} {\bibfnamefont {T.}~\bibnamefont {Olsen}},\ }\href
  {https://doi.org/10.1088/1361-648X/acab4b} {\bibfield  {journal} {\bibinfo
  {journal} {J. Phys. Condens. Matter}\ }\textbf {\bibinfo {volume} {35}},\
  \bibinfo {pages} {105802} (\bibinfo {year} {2023})}\BibitemShut {NoStop}%
\bibitem [{\citenamefont {Skovhus}\ \emph {et~al.}(2025)\citenamefont
  {Skovhus}, \citenamefont {Pavizhakumari},\ and\ \citenamefont
  {Olsen}}]{Skovhus2025}%
  \BibitemOpen
  \bibfield  {author} {\bibinfo {author} {\bibfnamefont {T.}~\bibnamefont
  {Skovhus}}, \bibinfo {author} {\bibfnamefont {V.~R.}\ \bibnamefont
  {Pavizhakumari}},\ and\ \bibinfo {author} {\bibfnamefont {T.}~\bibnamefont
  {Olsen}},\ }\href {https://doi.org/10.1103/nsfz-gpzn} {\bibfield  {journal}
  {\bibinfo  {journal} {Physical Review B}\ }\textbf {\bibinfo {volume}
  {112}},\ \bibinfo {pages} {205127} (\bibinfo {year} {2025})}\BibitemShut
  {NoStop}%
\bibitem [{\citenamefont {Wolf}\ \emph {et~al.}(2001)\citenamefont {Wolf},
  \citenamefont {Awschalom}, \citenamefont {Buhrman}, \citenamefont {Daughton},
  \citenamefont {von Moln\'{a}r}, \citenamefont {Roukes}, \citenamefont
  {Chtchelkanova},\ and\ \citenamefont {Treger}}]{Wolf2001}%
  \BibitemOpen
  \bibfield  {author} {\bibinfo {author} {\bibfnamefont {S.~A.}\ \bibnamefont
  {Wolf}}, \bibinfo {author} {\bibfnamefont {D.~D.}\ \bibnamefont {Awschalom}},
  \bibinfo {author} {\bibfnamefont {R.~A.}\ \bibnamefont {Buhrman}}, \bibinfo
  {author} {\bibfnamefont {J.~M.}\ \bibnamefont {Daughton}}, \bibinfo {author}
  {\bibfnamefont {S.}~\bibnamefont {von Moln\'{a}r}}, \bibinfo {author}
  {\bibfnamefont {M.~L.}\ \bibnamefont {Roukes}}, \bibinfo {author}
  {\bibfnamefont {A.~Y.}\ \bibnamefont {Chtchelkanova}},\ and\ \bibinfo
  {author} {\bibfnamefont {D.~M.}\ \bibnamefont {Treger}},\ }\href
  {https://doi.org/10.1126/science.1065389} {\bibfield  {journal} {\bibinfo
  {journal} {Science}\ }\textbf {\bibinfo {volume} {294}},\ \bibinfo {pages}
  {1488} (\bibinfo {year} {2001})}\BibitemShut {NoStop}%
\bibitem [{\citenamefont {Neusser}\ and\ \citenamefont
  {Grundler}(2009)}]{Neusser2009}%
  \BibitemOpen
  \bibfield  {author} {\bibinfo {author} {\bibfnamefont {S.}~\bibnamefont
  {Neusser}}\ and\ \bibinfo {author} {\bibfnamefont {D.}~\bibnamefont
  {Grundler}},\ }\href {https://doi.org/10.1002/adma.200900809} {\bibfield
  {journal} {\bibinfo  {journal} {Advanced Materials}\ }\textbf {\bibinfo
  {volume} {21}},\ \bibinfo {pages} {2927} (\bibinfo {year}
  {2009})}\BibitemShut {NoStop}%
\bibitem [{\citenamefont {Lenk}\ \emph {et~al.}(2011)\citenamefont {Lenk},
  \citenamefont {Ulrichs}, \citenamefont {Garbs},\ and\ \citenamefont
  {M\"{u}nzenberg}}]{Lenk2011}%
  \BibitemOpen
  \bibfield  {author} {\bibinfo {author} {\bibfnamefont {B.}~\bibnamefont
  {Lenk}}, \bibinfo {author} {\bibfnamefont {H.}~\bibnamefont {Ulrichs}},
  \bibinfo {author} {\bibfnamefont {F.}~\bibnamefont {Garbs}},\ and\ \bibinfo
  {author} {\bibfnamefont {M.}~\bibnamefont {M\"{u}nzenberg}},\ }\href
  {https://doi.org/10.1016/j.physrep.2011.06.003} {\bibfield  {journal}
  {\bibinfo  {journal} {Physics Reports}\ }\textbf {\bibinfo {volume} {507}},\
  \bibinfo {pages} {107} (\bibinfo {year} {2011})}\BibitemShut {NoStop}%
\bibitem [{\citenamefont {Chumak}\ \emph {et~al.}(2015)\citenamefont {Chumak},
  \citenamefont {Vasyuchka}, \citenamefont {Serga},\ and\ \citenamefont
  {Hillebrands}}]{Chumak2015}%
  \BibitemOpen
  \bibfield  {author} {\bibinfo {author} {\bibfnamefont {A.}~\bibnamefont
  {Chumak}}, \bibinfo {author} {\bibfnamefont {V.}~\bibnamefont {Vasyuchka}},
  \bibinfo {author} {\bibfnamefont {A.}~\bibnamefont {Serga}},\ and\ \bibinfo
  {author} {\bibfnamefont {B.}~\bibnamefont {Hillebrands}},\ }\href
  {https://doi.org/10.1038/nphys3347} {\bibfield  {journal} {\bibinfo
  {journal} {Nature Physics}\ }\textbf {\bibinfo {volume} {11}},\ \bibinfo
  {pages} {453} (\bibinfo {year} {2015})}\BibitemShut {NoStop}%
\bibitem [{\citenamefont {Qin}\ \emph {et~al.}(2015)\citenamefont {Qin},
  \citenamefont {Zakeri}, \citenamefont {Ernst}, \citenamefont {Sandratskii},
  \citenamefont {Buczek}, \citenamefont {Marmodoro}, \citenamefont {Chuang},
  \citenamefont {Zhang},\ and\ \citenamefont {Kirschner}}]{Qin2015}%
  \BibitemOpen
  \bibfield  {author} {\bibinfo {author} {\bibfnamefont {H.~J.}\ \bibnamefont
  {Qin}}, \bibinfo {author} {\bibfnamefont {K.}~\bibnamefont {Zakeri}},
  \bibinfo {author} {\bibfnamefont {A.}~\bibnamefont {Ernst}}, \bibinfo
  {author} {\bibfnamefont {L.~M.}\ \bibnamefont {Sandratskii}}, \bibinfo
  {author} {\bibfnamefont {P.}~\bibnamefont {Buczek}}, \bibinfo {author}
  {\bibfnamefont {A.}~\bibnamefont {Marmodoro}}, \bibinfo {author}
  {\bibfnamefont {T.~H.}\ \bibnamefont {Chuang}}, \bibinfo {author}
  {\bibfnamefont {Y.}~\bibnamefont {Zhang}},\ and\ \bibinfo {author}
  {\bibfnamefont {J.}~\bibnamefont {Kirschner}},\ }\href
  {https://doi.org/10.1038/ncomms7126} {\bibfield  {journal} {\bibinfo
  {journal} {Nature Communications}\ }\textbf {\bibinfo {volume} {6}},\
  \bibinfo {pages} {6126} (\bibinfo {year} {2015})}\BibitemShut {NoStop}%
\bibitem [{\citenamefont {Schoen}\ \emph {et~al.}(2016)\citenamefont {Schoen},
  \citenamefont {Thonig}, \citenamefont {Schneider}, \citenamefont {Silva},
  \citenamefont {Nembach}, \citenamefont {Eriksson}, \citenamefont {Karis},\
  and\ \citenamefont {Shaw}}]{Schoen2016}%
  \BibitemOpen
  \bibfield  {author} {\bibinfo {author} {\bibfnamefont {M.~A.}\ \bibnamefont
  {Schoen}}, \bibinfo {author} {\bibfnamefont {D.}~\bibnamefont {Thonig}},
  \bibinfo {author} {\bibfnamefont {M.~L.}\ \bibnamefont {Schneider}}, \bibinfo
  {author} {\bibfnamefont {T.~J.}\ \bibnamefont {Silva}}, \bibinfo {author}
  {\bibfnamefont {H.~T.}\ \bibnamefont {Nembach}}, \bibinfo {author}
  {\bibfnamefont {O.}~\bibnamefont {Eriksson}}, \bibinfo {author}
  {\bibfnamefont {O.}~\bibnamefont {Karis}},\ and\ \bibinfo {author}
  {\bibfnamefont {J.~M.}\ \bibnamefont {Shaw}},\ }\href
  {https://doi.org/10.1038/nphys3770} {\bibfield  {journal} {\bibinfo
  {journal} {Nature Physics}\ }\textbf {\bibinfo {volume} {12}},\ \bibinfo
  {pages} {839} (\bibinfo {year} {2016})}\BibitemShut {NoStop}%
\bibitem [{\citenamefont {Paischer}\ \emph {et~al.}(2024)\citenamefont
  {Paischer}, \citenamefont {Eilmsteiner}, \citenamefont {Maznichenko},
  \citenamefont {Buczek}, \citenamefont {Zakeri}, \citenamefont {Ernst},\ and\
  \citenamefont {Buczek}}]{Paischer2024}%
  \BibitemOpen
  \bibfield  {author} {\bibinfo {author} {\bibfnamefont {S.}~\bibnamefont
  {Paischer}}, \bibinfo {author} {\bibfnamefont {D.}~\bibnamefont
  {Eilmsteiner}}, \bibinfo {author} {\bibfnamefont {I.}~\bibnamefont
  {Maznichenko}}, \bibinfo {author} {\bibfnamefont {N.}~\bibnamefont {Buczek}},
  \bibinfo {author} {\bibfnamefont {K.}~\bibnamefont {Zakeri}}, \bibinfo
  {author} {\bibfnamefont {A.}~\bibnamefont {Ernst}},\ and\ \bibinfo {author}
  {\bibfnamefont {P.~A.}\ \bibnamefont {Buczek}},\ }\bibfield  {journal}
  {\bibinfo  {journal} {Physical Review B}\ }\textbf {\bibinfo {volume}
  {109}},\ \href {https://doi.org/10.1103/PhysRevB.109.L220405}
  {10.1103/PhysRevB.109.L220405} (\bibinfo {year} {2024})\BibitemShut {NoStop}%
\bibitem [{\citenamefont {Paischer}\ \emph {et~al.}(2021)\citenamefont
  {Paischer}, \citenamefont {Buczek}, \citenamefont {Buczek}, \citenamefont
  {Eilmsteiner},\ and\ \citenamefont {Ernst}}]{Paischer2021}%
  \BibitemOpen
  \bibfield  {author} {\bibinfo {author} {\bibfnamefont {S.}~\bibnamefont
  {Paischer}}, \bibinfo {author} {\bibfnamefont {P.~A.}\ \bibnamefont
  {Buczek}}, \bibinfo {author} {\bibfnamefont {N.}~\bibnamefont {Buczek}},
  \bibinfo {author} {\bibfnamefont {D.}~\bibnamefont {Eilmsteiner}},\ and\
  \bibinfo {author} {\bibfnamefont {A.}~\bibnamefont {Ernst}},\ }\href
  {https://doi.org/10.1103/PhysRevB.104.024403} {\bibfield  {journal} {\bibinfo
   {journal} {Physical Review B}\ }\textbf {\bibinfo {volume} {104}},\ \bibinfo
  {pages} {024403} (\bibinfo {year} {2021})}\BibitemShut {NoStop}%
\bibitem [{\citenamefont {Pavizhakumari}\ \emph {et~al.}(2025)\citenamefont
  {Pavizhakumari}, \citenamefont {Skovhus},\ and\ \citenamefont
  {Olsen}}]{RajeevPavizhakumari2025}%
  \BibitemOpen
  \bibfield  {author} {\bibinfo {author} {\bibfnamefont {V.~R.}\ \bibnamefont
  {Pavizhakumari}}, \bibinfo {author} {\bibfnamefont {T.}~\bibnamefont
  {Skovhus}},\ and\ \bibinfo {author} {\bibfnamefont {T.}~\bibnamefont
  {Olsen}},\ }\href {https://doi.org/10.1088/1361-648X/ada65c} {\bibfield
  {journal} {\bibinfo  {journal} {Journal of Physics: Condensed Matter}\
  }\textbf {\bibinfo {volume} {37}},\ \bibinfo {pages} {115806} (\bibinfo
  {year} {2025})}\BibitemShut {NoStop}%
\bibitem [{\citenamefont {Beaurepaire}\ \emph {et~al.}(1996)\citenamefont
  {Beaurepaire}, \citenamefont {Merle}, \citenamefont {Daunois},\ and\
  \citenamefont {Bigot}}]{Beaurepaire1996}%
  \BibitemOpen
  \bibfield  {author} {\bibinfo {author} {\bibfnamefont {E.}~\bibnamefont
  {Beaurepaire}}, \bibinfo {author} {\bibfnamefont {J.-C.}\ \bibnamefont
  {Merle}}, \bibinfo {author} {\bibfnamefont {A.}~\bibnamefont {Daunois}},\
  and\ \bibinfo {author} {\bibfnamefont {J.-Y.}\ \bibnamefont {Bigot}},\ }\href
  {https://doi.org/10.1103/PhysRevLett.76.4250} {\bibfield  {journal} {\bibinfo
   {journal} {Physical Review Letters}\ }\textbf {\bibinfo {volume} {76}},\
  \bibinfo {pages} {4250} (\bibinfo {year} {1996})}\BibitemShut {NoStop}%
\bibitem [{\citenamefont {Koopmans}\ \emph {et~al.}(2005)\citenamefont
  {Koopmans}, \citenamefont {Ruigrok}, \citenamefont {Longa},\ and\
  \citenamefont {de~Jonge}}]{Koopmans2005}%
  \BibitemOpen
  \bibfield  {author} {\bibinfo {author} {\bibfnamefont {B.}~\bibnamefont
  {Koopmans}}, \bibinfo {author} {\bibfnamefont {J.~J.~M.}\ \bibnamefont
  {Ruigrok}}, \bibinfo {author} {\bibfnamefont {F.~D.}\ \bibnamefont {Longa}},\
  and\ \bibinfo {author} {\bibfnamefont {W.~J.~M.}\ \bibnamefont {de~Jonge}},\
  }\href {https://doi.org/10.1103/PhysRevLett.95.267207} {\bibfield  {journal}
  {\bibinfo  {journal} {Physical Review Letters}\ }\textbf {\bibinfo {volume}
  {95}},\ \bibinfo {pages} {267207} (\bibinfo {year} {2005})}\BibitemShut
  {NoStop}%
\bibitem [{\citenamefont {Kirilyuk}\ \emph {et~al.}(2010)\citenamefont
  {Kirilyuk}, \citenamefont {Kimel},\ and\ \citenamefont
  {Rasing}}]{Kirilyuk2010}%
  \BibitemOpen
  \bibfield  {author} {\bibinfo {author} {\bibfnamefont {A.}~\bibnamefont
  {Kirilyuk}}, \bibinfo {author} {\bibfnamefont {A.~V.}\ \bibnamefont
  {Kimel}},\ and\ \bibinfo {author} {\bibfnamefont {T.}~\bibnamefont
  {Rasing}},\ }\href {https://doi.org/10.1103/RevModPhys.82.2731} {\bibfield
  {journal} {\bibinfo  {journal} {Reviews of Modern Physics}\ }\textbf
  {\bibinfo {volume} {82}},\ \bibinfo {pages} {2731} (\bibinfo {year}
  {2010})}\BibitemShut {NoStop}%
\bibitem [{\citenamefont {Pankratova}\ \emph {et~al.}(2022)\citenamefont
  {Pankratova}, \citenamefont {Miranda}, \citenamefont {Thonig}, \citenamefont
  {Pereiro}, \citenamefont {Sj\"{o}qvist}, \citenamefont {Delin}, \citenamefont
  {Eriksson},\ and\ \citenamefont {Bergman}}]{Pankratova2022}%
  \BibitemOpen
  \bibfield  {author} {\bibinfo {author} {\bibfnamefont {M.}~\bibnamefont
  {Pankratova}}, \bibinfo {author} {\bibfnamefont {I.~P.}\ \bibnamefont
  {Miranda}}, \bibinfo {author} {\bibfnamefont {D.}~\bibnamefont {Thonig}},
  \bibinfo {author} {\bibfnamefont {M.}~\bibnamefont {Pereiro}}, \bibinfo
  {author} {\bibfnamefont {E.}~\bibnamefont {Sj\"{o}qvist}}, \bibinfo {author}
  {\bibfnamefont {A.}~\bibnamefont {Delin}}, \bibinfo {author} {\bibfnamefont
  {O.}~\bibnamefont {Eriksson}},\ and\ \bibinfo {author} {\bibfnamefont
  {A.}~\bibnamefont {Bergman}},\ }\href
  {https://doi.org/10.1103/PhysRevB.106.174407} {\bibfield  {journal} {\bibinfo
   {journal} {Physical Review B}\ }\textbf {\bibinfo {volume} {106}},\ \bibinfo
  {pages} {174407} (\bibinfo {year} {2022})}\BibitemShut {NoStop}%
\bibitem [{\citenamefont {Nabok}\ \emph {et~al.}(2021)\citenamefont {Nabok},
  \citenamefont {Bl{\"{u}}gel},\ and\ \citenamefont {Friedrich}}]{Nabok2021}%
  \BibitemOpen
  \bibfield  {author} {\bibinfo {author} {\bibfnamefont {D.}~\bibnamefont
  {Nabok}}, \bibinfo {author} {\bibfnamefont {S.}~\bibnamefont
  {Bl{\"{u}}gel}},\ and\ \bibinfo {author} {\bibfnamefont {C.}~\bibnamefont
  {Friedrich}},\ }\href {https://doi.org/10.1038/s41524-021-00649-8} {\bibfield
   {journal} {\bibinfo  {journal} {npj Computational Materials}\ }\textbf
  {\bibinfo {volume} {7}},\ \bibinfo {pages} {178} (\bibinfo {year}
  {2021})}\BibitemShut {NoStop}%
\bibitem [{\citenamefont {Paischer}\ \emph {et~al.}(2023)\citenamefont
  {Paischer}, \citenamefont {Vignale}, \citenamefont {Katsnelson},
  \citenamefont {Ernst},\ and\ \citenamefont {Buczek}}]{Paischer2023}%
  \BibitemOpen
  \bibfield  {author} {\bibinfo {author} {\bibfnamefont {S.}~\bibnamefont
  {Paischer}}, \bibinfo {author} {\bibfnamefont {G.}~\bibnamefont {Vignale}},
  \bibinfo {author} {\bibfnamefont {M.~I.}\ \bibnamefont {Katsnelson}},
  \bibinfo {author} {\bibfnamefont {A.}~\bibnamefont {Ernst}},\ and\ \bibinfo
  {author} {\bibfnamefont {P.~A.}\ \bibnamefont {Buczek}},\ }\href
  {https://doi.org/10.1103/PhysRevB.107.134410} {\bibfield  {journal} {\bibinfo
   {journal} {Physical Review B}\ }\textbf {\bibinfo {volume} {107}},\ \bibinfo
  {pages} {134410} (\bibinfo {year} {2023})}\BibitemShut {NoStop}%
\bibitem [{\citenamefont {Skovhus}\ and\ \citenamefont
  {Thunstr\"{o}m}(2026)}]{meomdata}%
  \BibitemOpen
  \bibfield  {author} {\bibinfo {author} {\bibfnamefont {T.}~\bibnamefont
  {Skovhus}}\ and\ \bibinfo {author} {\bibfnamefont {P.}~\bibnamefont
  {Thunstr\"{o}m}},\ }\href {https://doi.org/10.57804/3mx0-zx05} {\bibinfo
  {title} {Magnon frequency and lifetime dispersion of a prototypical itinerant
  ferromagnet}},\ \bibinfo {howpublished} {Swedish National Data Service}
  (\bibinfo {year} {2026})\BibitemShut {NoStop}%
\bibitem [{\citenamefont {Niesert}(2011)}]{Niesert2011}%
  \BibitemOpen
  \bibfield  {author} {\bibinfo {author} {\bibfnamefont {M.}~\bibnamefont
  {Niesert}},\ }\emph {\bibinfo {title} {{Ab initio calculations of spin-wave
  spectra from time-dependent density-functional theory}}},\ \href@noop {}
  {Ph.D. thesis},\ \bibinfo  {school} {RWTH Aachen University} (\bibinfo {year}
  {2011})\BibitemShut {NoStop}%
\bibitem [{\citenamefont {Kubo}(1966)}]{Kubo1966}%
  \BibitemOpen
  \bibfield  {author} {\bibinfo {author} {\bibfnamefont {R.}~\bibnamefont
  {Kubo}},\ }\href
  {http://iopscience.iop.org/article/10.1088/0034-4885/29/1/306/meta}
  {\bibfield  {journal} {\bibinfo  {journal} {Rep. Prog. Phys.}\ }\textbf
  {\bibinfo {volume} {29}},\ \bibinfo {pages} {255} (\bibinfo {year}
  {1966})}\BibitemShut {NoStop}%
\bibitem [{\citenamefont {Jensen}\ and\ \citenamefont
  {Mackintosh}(1991)}]{Jensen1991}%
  \BibitemOpen
  \bibfield  {author} {\bibinfo {author} {\bibfnamefont {J.}~\bibnamefont
  {Jensen}}\ and\ \bibinfo {author} {\bibfnamefont {A.~R.}\ \bibnamefont
  {Mackintosh}},\ }\href@noop {} {\emph {\bibinfo {title} {Rare Earth
  Magnetism: Structures and excitations}}},\ The International Series of
  Monographs on Physics\ (\bibinfo  {publisher} {Clarendon Press, Oxford},\
  \bibinfo {year} {1991})\BibitemShut {NoStop}%
\end{thebibliography}%

\appendix
\onecolumngrid
\begin{center}
    \textbf{End Matter}
\end{center}
\twocolumngrid

Our model for a prototypical itinerant ferromagnet is constructed to account for two types of contributions to the magnon dynamics: contributions from pairs of spin-polarized bands that reside fully below/above the Fermi level (typically narrow), and contributions from pairs of spin-polarized bands penetrating the Fermi level (typically dispersive). Phenomenologically, one may separate these two contributions in the noninteracting susceptibility of a given single-particle band structure (tight-binding, Hartree-Fock, Kohn-Sham etc.),
\begin{equation}
    \chi_\mathrm{KS}^{+-}(\mathbf{q},\omega)=\chi_\mathrm{KS,loc}^{+-}(\mathbf{q},\omega) + \chi_\mathrm{KS,deloc}^{+-}(\mathbf{q},\omega),
    \label{eq:chi kohn-sham}
\end{equation}
and model them separately.
\begin{figure*}[tb]
    \centering
    \includegraphics[scale=1.0]{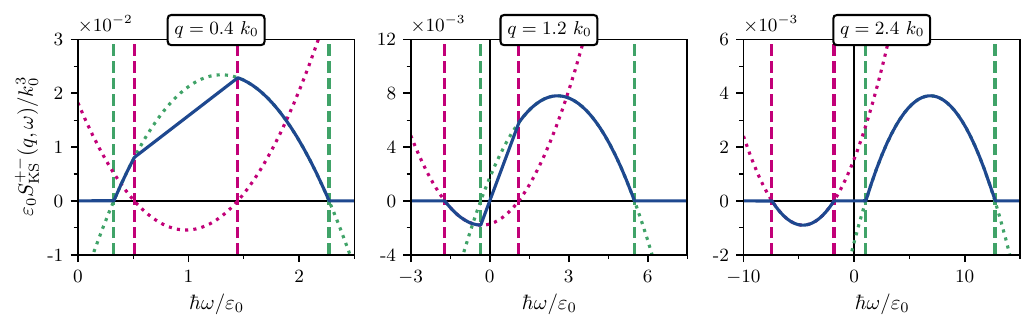}
    \caption{Scattering function (blue) of the noninteracting spin-polarized homogeneous electron gas at relative spin-polarization $\zeta=0.8$, composed of parabolic contributions $3\sigma n_\sigma\big(w_\sigma(q)^2-[\hbar z - \varepsilon_\sigma^p(q)]^2\big)/(4 w_\sigma(q)^3)$ (dotted lines) from the majority (green) and minority (magenta) Fermi surfaces cut at the roots $\varepsilon_\sigma^p(q) \pm w_\sigma(q)$ of each parabola (vertical dashed lines). The three panels (left to right) represent the three qualitatively different regions $q<k_{\mathrm{F}\uparrow}-k_{\mathrm{F}\downarrow}$, $k_{\mathrm{F}\uparrow}-k_{\mathrm{F}\downarrow}<q<k_{\mathrm{F}\uparrow}+k_{\mathrm{F}\downarrow}$, and $q>k_{\mathrm{F}\uparrow}+k_{\mathrm{F}\downarrow}$. All quantities are given in units of the paramagnetic Fermi wave vector $k_0^3=3\pi^2n$ and the paramagnetic Fermi level $\varepsilon_0=\varepsilon(k_0)$.}
    \label{fig:heg_stoner_regions}
\end{figure*}
In this work, we model $\chi_\mathrm{KS,deloc}^{+-}$ using the noninteracting susceptibility of the spin-polarized homogeneous electron gas (HEG) \cite{Moriya1985,Niesert2011,Friedrich2020} and construct $\chi_\mathrm{KS,loc}^{+-}$ with a narrow spectral weight distribution centered at the exchange splitting $\Delta_\mathrm{x}$. The many-body susceptibility of the Goldstone mode is then calculated by inverting a Dyson equation (at the level of RPA/ALDA),
\begin{equation}
    \chi^{+-}(\mathbf{q},\omega) =\chi_\mathrm{KS}^{+-}(\mathbf{q},\omega) - \chi_\mathrm{KS}^{+-}(\mathbf{q},\omega)\frac{\Delta_\mathrm{x}}{n^z} \chi^{+-}(\mathbf{q},\omega),
    \label{eq:RPA Dyson}
\end{equation}
where $n^z$ is the total spectral weight of $\chi_\mathrm{KS}^{+-}(\mathbf{q},\omega)$, which is directly related to the ground-state spin-polarization thanks to a sum rule \cite{Kubo1966,Jensen1991,Skovhus2021}.

The single-particle (Kohn-Sham) susceptibility of the spin-polarized homogeneous electron gas, see also Eq. \eqref{eq:mode projected susceptibility}, is for complex frequencies $z$ in the upper-half complex frequency plane given by the Lindhard function
\begin{equation}
    \chi_\mathrm{KS}^{+-}(\mathbf{q},z) = \frac{1}{V} \sum_\mathbf{k} \frac{f_{\mathbf{k}\uparrow} - f_{\mathbf{k}+\mathbf{q}\downarrow}}{\hbar z - (\varepsilon_{\mathbf{k}+\mathbf{q}\downarrow} - \varepsilon_{\mathbf{k}\uparrow})},
    \label{eq:lindhard function}
\end{equation}
where $\varepsilon_{\mathbf{k}\sigma}=\varepsilon_\sigma(|\mathbf{k}|)$ are the band energies, $f_{\mathbf{k}\sigma}=\theta(k_{\mathrm{F}\sigma}-|\mathbf{k}|)$ the band occupations, and $V$ the gas volume. The two parabolic bands are split by exchange, $\varepsilon_\sigma(k)=\varepsilon(k) - \sigma \Delta_\mathrm{x}/2$, where $\varepsilon(k) = \hbar^2k^2/(2m)$, and for a gas with relative spin-polarization $\zeta=(n_\uparrow-n_\downarrow)/n$, the exchange splitting can be inferred from the Fermi wave vectors, $\Delta_\mathrm{x}=\varepsilon(k_{\mathrm{F}\uparrow}) - \varepsilon(k_{\mathrm{F}\downarrow})$ with $k_{\mathrm{F}\sigma}^3=6\pi^2n_\sigma=3\pi^2n(1+\sigma\zeta)$. The reciprocal space integral \eqref{eq:lindhard function} can be solved exactly,
\begin{align}
    \chi_\mathrm{KS}^{+-}(q,z) &= \sum_\sigma \frac{3 \sigma n_\sigma}{2 w_\sigma(q)^3}
    \bigg(
        w_\sigma(q) [\hbar z - \varepsilon_\sigma^p(q)] 
        \nonumber \\
        &+\frac{w_\sigma(q)^2-[\hbar z - \varepsilon_\sigma^p(q)]^2}{2}\chi_{w_\sigma(q)}[\hbar z - \varepsilon_\sigma^p(q)]
    \bigg),
    \label{eq:HEG KS susceptibility}
\end{align}
where $w_\sigma(q)=\varepsilon(\sqrt{2k_{\mathrm{F}\sigma}q})$, $\varepsilon_\sigma^p(q)=\Delta_\mathrm{x}+\sigma\varepsilon(q)$, and
\begin{equation}
    \chi_w(\varepsilon)=\ln\frac{\varepsilon + w}{\varepsilon - w}
\end{equation}
denotes the complex rectangular function of half-width $w>0$. The sum over spin in Eq. \eqref{eq:HEG KS susceptibility} accounts for contributions to the Lindhard function \eqref{eq:lindhard function} from $f_{\mathbf{k}\uparrow}$ and $f_{\mathbf{k}+\mathbf{q}\downarrow}$ respectively. 
For frequencies on the real axis,
\begin{equation}
    \lim_{\eta\rightarrow 0^+} \chi_w(\hbar\omega+i\hbar\eta) = \ln \left|\frac{\hbar\omega + w}{\hbar\omega - w}\right|-i\pi\theta(w-|\hbar\omega|),
\end{equation}
meaning that both Fermi surfaces supply the scattering function $S_\mathrm{KS}^{+-}(q,\omega)=-\mathrm{Im}\,\chi_\mathrm{KS}^{+-}(q,\omega)/\pi$ with spectral weight shaped as a cut parabola of half-width $w_\sigma(q)$ centered at $\varepsilon_\sigma^p(q)$, see Fig. \ref{fig:heg_stoner_regions}. At the cuts/bounds of each parabola (its roots), $\chi_\mathrm{KS}^{+-}(q,\omega)$ exhibits a derivative discontinuity, which in turn induces nonanalytical effects in the magnon dispersion. In particular, the inverse magnon lifetime $-\mathrm{Im}\,z(q)$ goes from zero to a linear function of $q$ at the lower majority bound $\varepsilon_\uparrow^p(q) - w_\uparrow(q)$, while the lower minority bound at $\varepsilon_\downarrow^p(q) - w_\downarrow(q)$ induces a discontinuity in both the global lineshape maximum and magnon resonance frequency.

\begin{figure*}[tb]
    \centering
    \includegraphics[scale=1.0]{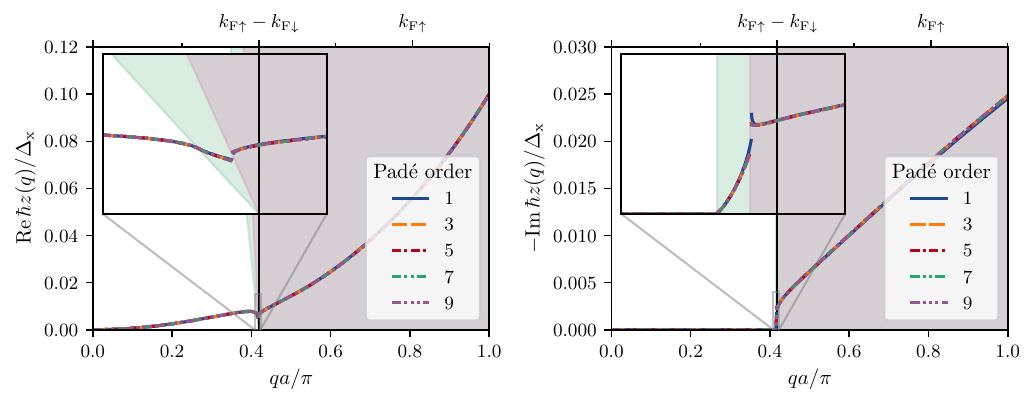}
    \caption{Real (left) and imaginary (right) parts of the magnon resonance frequency calculated by self-consistently solving Eqs. \eqref{eq:meom complex frequency solution} and \eqref{eq:linear interpolation omega} using Pad\'{e} approximants \eqref{eq:pade approximant} of odd order $N=2n+1$.}
    \label{fig:ifm_magnon_resonances}
\end{figure*}
In order to model $\chi_\mathrm{KS,loc}^{+-}$ for a prototypical itinerant ferromagnet, we note that the corresponding Lindhard function \eqref{eq:lindhard function} only involves transitions from majority band(s) to minority band(s). Inspired by the HEG Lindhard function, we model the spectral weight of these transitions with a single cut parabola of half-width $w(q)=E_\mathrm{b}\sin\left(q a/2\right)$ centered at the exchange splitting,
\begin{align}
    \chi_\mathrm{KS,\mathrm{loc}}^{+-}(q,z) = &\frac{3 n_\mathrm{loc}^z}{2 w(q)^3}
    \bigg(
        w(q) [\hbar z - \Delta_\mathrm{x}] 
        \nonumber \\
        &+\frac{w(q)^2-[\hbar z - \Delta_\mathrm{x}]^2}{2}\chi_{w(q)}[\hbar z - \Delta_\mathrm{x}]
    \bigg),
    \label{eq:localized ks susceptibility}
\end{align}
where $E_\mathrm{b}<\Delta_\mathrm{x}$ represents (qualitatively) the width of the fully spin-polarized bands along the given quasi-crystal direction, $q\in[0,\pi/a]$. It is important to emphasize that the actual spectral shape of $\chi_\mathrm{KS,\mathrm{loc}}^{+-}(q,z)$ only has a minor quantitative influence on the magnon dispersion. As long as $E_\mathrm{b}\ll\Delta_\mathrm{x}$, all qualitative features remain unchanged if $\chi_\mathrm{KS,\mathrm{loc}}^{+-}(q,z)$ is replaced with e.g. a rectangular spectral distribution. 
For the present model, we fix the pseudo lattice constant $a$ based on the Fermi surface radii $a=\pi(k_{\mathrm{F}\downarrow}+k_{\mathrm{F}\uparrow}/2)^{-1}$, set the relative spin-polarization of $\chi_\mathrm{KS,\mathrm{deloc}}^{+-}(q,z)$ to $\zeta=0.8$, use an effective band width of $E_\mathrm{b}=0.2\, \Delta_\mathrm{x}$ for $\chi_\mathrm{KS,\mathrm{loc}}^{+-}(q,z)$ and make the fully spin-polarized bands responsible for 60\% of the total spin-polarization $n^z=n_\mathrm{loc}^z + n_\mathrm{deloc}^z=5n_\mathrm{deloc}^z/2$.

Inverting the Dyson equation \eqref{eq:RPA Dyson} using the combined analytical models \eqref{eq:HEG KS susceptibility} and \eqref{eq:localized ks susceptibility}, the inverse susceptibility $\chi^{-1}(q,z)=1/[2\chi^{+-}(q,z)]$ and all its frequency derivatives can be computed analytically in the upper-half frequency plane. In order to continue $\chi^{-1}(q,z)$ to $\mathrm{Im}\,z<0$, it is important to note that $\chi^{+-}(q,\omega)\propto1/(\omega-z_\mathrm{eff})$ for frequencies $\omega$ far below/above the bounds of the Stoner continuum. This means that $\chi^{-1}(q,z)$ has linear asymptotes and can be continued efficiently using Pad\'{e} approximants,
\begin{equation}
    \chi_{(n+1)/n}^{-1}(q,z)=\frac{\sum_{j=0}^{n+1} a_j(q,\mathrm{Re}\,z)(i\,\mathrm{Im}\,z)^j}{1+\sum_{k=1}^n b_k(q,\mathrm{Re}\,z)(i\,\mathrm{Im}\,z)^j},\quad n\in\mathbb{N}_0
    \label{eq:pade approximant}
\end{equation}
of odd order $N=2n+1$. The coefficients $a_j(q,\mathrm{Re}\,z)$ and $b_k(q,\mathrm{Re}\,z)$ are determined such that the $N$'th order approximant \eqref{eq:pade approximant} reproduces exactly the first $N+1$ terms of $\chi^{-1}(q,\omega)$'s Taylor series expansion around the real frequency $\omega=\mathrm{Re}\,z$. Doing so, we can robustly compute the inverse susceptibility roots $z(q)$, both on the real axis and in the lower complex frequency plane. In Fig. \ref{fig:ifm_magnon_resonances} we present the convergence of the determined magnon resonances as a function of the Pad\'{e} order $N$. Because the roots lie relatively close to the real frequency axis, the Pad\'{e} approximant series \eqref{eq:pade approximant} converges extremely fast. Already at lowest order $N=1$, one has a good estimate of the magnon resonance frequency. Essentially, this means that one could, to a good approximation, live with a linear extrapolation of $\chi^{-1}(q,z)$ to the lower-half complex frequency plane.

\end{document}